%% file: main.tex
\documentclass[conference]{IEEEtran}
\usepackage{cite}
\usepackage{amsmath,amssymb,amsfonts}
\usepackage{algorithm}
\usepackage{algorithmic}
\usepackage{comment}
\usepackage{lscape}
\usepackage{multirow}
\usepackage{graphicx}
\usepackage{textcomp}
\usepackage{xcolor}
\usepackage[hyphens]{url}
\usepackage[normalem]{ulem}

\floatname{algorithm}{Algorithm}

\newcommand{\srt}[1]{\textcolor{black}{#1}}

\newcommand{\newcontent}[1]{\textcolor{black}{#1}}
\newcommand{\rewritedesign}[1]{\textcolor{black}{#1}}
\newcommand{\rewriteresults}[1]{\textcolor{black}{#1}}

\def\BibTeX{{\rm B\kern-.05em{\sc i\kern-.025em b}\kern-.08em
    T\kern-.1667em\lower.7ex\hbox{E}\kern-.125emX}}

\title{
    \textit{MAD}-Max Beyond Single-Node: 
    Enabling Large Machine Learning \textit{M}odel \textit{A}cceleration on \textit{D}istributed Systems}
\author{
    Samuel Hsia$^{1,2}$, Alicia Golden$^{1,2}$, Bilge Acun$^{1}$, Newsha Ardalani$^{1}$, Zachary DeVito$^{1}$, 
    \\Gu-Yeon Wei$^{2}$, David Brooks$^{2}$, Carole-Jean Wu$^{1}$\\ \\
    $^{1}$FAIR at Meta, $^{2}$Harvard University\\ \\
    shsia@g.harvard.edu, carolejeanwu@meta.com}

\begin{document}
\maketitle
\thispagestyle{plain}
\pagestyle{plain}


\begin{abstract}
\input{text_CR/abstract}
\end{abstract}

\input{text_CR/introduction}
\input{text_CR/background}
\input{text_CR/characterization}
\input{text_CR/design}
\input{text_CR/methodology}
\input{text_CR/results}

\input{text_CR/related_work_and_discussion}

\input{text_CR/conclusion}
\input{text_CR/acknowledgements}


\bibliographystyle{IEEEtranS}
\bibliography{references}

\end{document}

%% file: text_CR/abstract.tex
Training and deploying large-scale machine learning models is time-consuming, requires significant distributed computing infrastructures, and incurs high operational costs.
Our analysis, grounded in real-world large model training on datacenter-scale infrastructures, reveals that 14$\sim$32\% of all GPU hours are spent on communication with no overlapping computation.  
To minimize this outstanding communication latency and other inherent at-scale inefficiencies, we introduce an agile performance modeling framework, MAD-Max.
This framework is designed to optimize parallelization strategies and facilitate hardware-software co-design opportunities.
Through the application of MAD-Max to a suite of real-world large-scale ML models on state-of-the-art GPU clusters, we showcase potential throughput enhancements of up to 2.24$\times$ for \textit{pre-training} and up to 5.27$\times$ for \textit{inference} scenarios, respectively.

%% file: text_CR/introduction.tex
\section{Introduction}~\label{sec:introduction}
Billion-parameter large language models (LLMs)~\cite{brown2020gpt3, 
openai2023gpt4, touvron2023llama, touvron2023llama2} power applications that have shown far-reaching impact across different domains~\cite{google2023bard, gozalobrizuela2023chatgpt, gozalobrizuela2023survey, openai2023chatgpt}.
Similarly, trillion-parameter recommendation models~\cite{mudigere2021zionex, zhang2022dhen} have demonstrated state-of-the-art user modeling and content understanding across search~\cite{anil2022factoryfloor, cheng2016wnd, jouppi2017tpu, zhao2019mtwnd}, social media~\cite{acun2020understanding, gupta2020architectural, hazelwood2018applied, yiSysml18}, e-commerce~\cite{zhou2019dien, zhou2018din}, and entertainment~\cite{he2017ncf}.
As these large-scale ML models increase in size and complexity~\cite{gupta2020architectural, hazelwood2018applied}, the corresponding training and inference workloads become ever more resource-intensive.
Without efficient mappings between these large-scale ML workloads and their underlying distributed systems, model training and exploration can easily require millions of GPU hours, levying high operational costs, compute resource requirements, and energy consumption~\cite{brown2020gpt3, touvron2023llama, touvron2023llama2}.

\begin{figure}[t!]
    \centering
    \includegraphics[width=\linewidth]{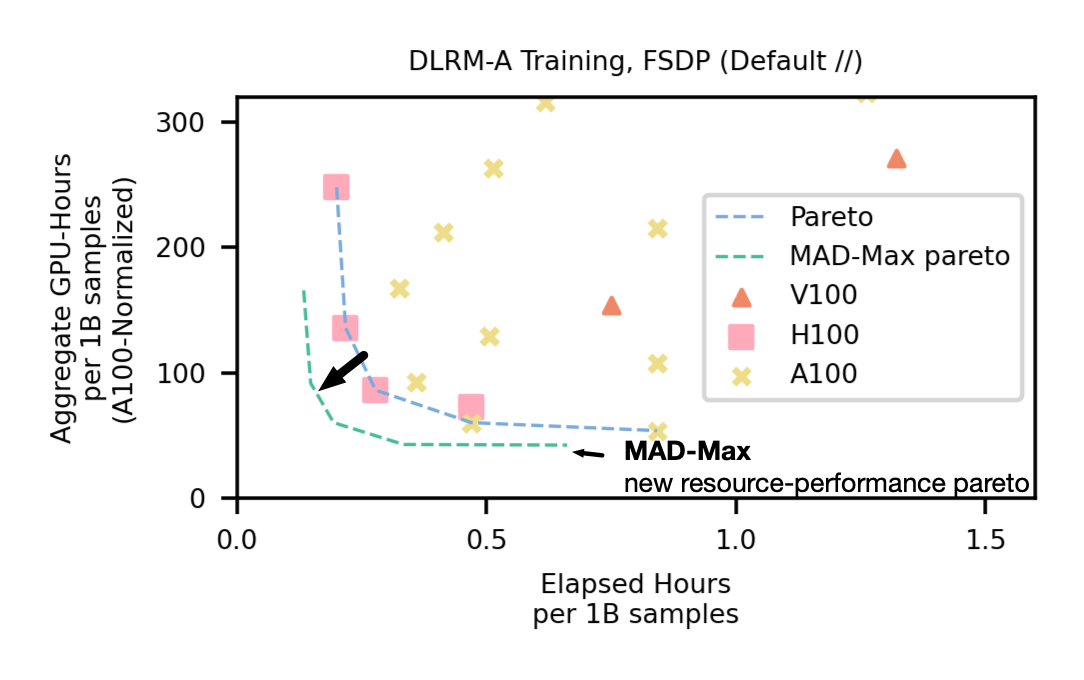}
    \caption{\srt{Our performance model -- MAD-Max -- improves upon the resource-performance pareto frontier of large-scale ML workloads by identifying new hardware-software mappings and solutions.}}
    \label{fig:motivation}
\end{figure}

\srt{Figure~\ref{fig:motivation} shows the projected resource-performance pareto frontier of training a state-of-the-art deep learning recommendation model (DLRM) using default workload-system mapping strategy on public cloud instances.
In this case, we quantify compute resource requirements with aggregate GPU hours per 1 billion samples, where aggregate GPU hours of different generations of GPUs are normalized based on the A100's peak FLOPS.}
Further improving upon this resource-performance pareto frontier requires researchers to take into account underlying distributed systems~\cite{firoozshahian2023mtia, jouppi2021tpu, jouppi2023tpu, jouppi2020tpu, nvidia2023h100, nvidia2018v100, nvidia2019dgx1, nvidia2023hgxh100} and how we map models and tasks onto underlying distributed systems -- \textit{parallelization strategy}.
In this paper, we propose a distributed ML performance model -- \textbf{\textit{MAD-Max}} -- for identifying potential avenues for improvement (green, dotted line).
Nonetheless, pinpointing the specific distributed systems and parallelization strategies needed for realizing these improvements in performance and operational compute resource requirements is challenging, as evidenced by the three general approaches for optimizing runtime performance of large ML models.

\srt{The first option involves applying industry-standard parallelization strategies (Figure~\ref{fig:motivation}: blue, dotted line) that target feasibility without fully optimizing hardware usage (e.g., FSDP~\cite{zhao2023fsdp, rajbhandari2020zero}).}
The second option is to custom-design custom hierarchical parallelization strategies specific to the model, task, and system~\cite{shoeybi2020megatronlm}. 
This maximizes hardware efficiency but is complex from an engineering standpoint and not easily adaptable across different tasks.
The third option is to use software tools to predict system performance before training or deployment, though existing tools have several limitations, such as being training-specific or hardware architecture-dependent.
To address the need for an agile exploration tool to identify parallelization strategies tailored to different use-cases, we introduce our distributed ML performance model -- \textit{MAD-Max} -- and evaluate it on a suite of real-world, large ML models, including deep learning recommender systems and LLMs~\cite{artetxe2022efficient, brown2020gpt3, chen2019alibaba, du2022glam, naumov2019dlrm, pei2019alibaba, touvron2023llama, touvron2023llama2, zhao2019mtwnd}.

In this work, we first characterize a suite of real-world, large ML models at both model- and datacenter-deployment scales (Section \ref{sec:characterization}).
At the model architecture level, we identify performance-critical hardware requirements based on the models' compute and memory characteristics.
At the datacenter scale, we quantify the required communication by conducting a fleet-wide training characterization, revealing that 14$\sim$32\% of all GPU hours are spent on \textit{communication with no concurrent computation} (i.e., exposed communication).

To enable agile exploration of the parallelization design space, MAD-Max first estimates the system performance of large-scale ML workloads.
The performance model takes in target ML model architecture, task details, parallelization scheme, and distributed system hardware to generated per-device traces.
These per-device traces are then pieced together to estimate the overall system performance of the target ML model and task.
Additionally, the performance model generates detailed breakdowns of both communication collectives and computation-communication overlap efficiency, enabling users to identify future optimization opportunities.
Our performance model is validated against multiple real-world large-scale distributed training experiments, demonstrating 97\% and 91\% performance prediction accuracies on serialized and overlapped execution, respectively.

Using MAD-Max, we identify parallelization strategies that result in throughput improvements across our suite of large ML models -- achieving up to 2.24$\times$ and 5.27$\times$ throughput improvements for pre-training and inference, respectively.
By extending our analysis to parallelization strategies that are not constrained by the memory capacities of existing training platforms, we discover strategies capable of delivering up to 2.43$\times$ and 12.13$\times$ throughput improvement for pre-training and inference, respectively.
Furthermore, MAD-Max provides critical insights on how model-level compute and communication requirements alter optimal parallelization strategy and increasing LLM context lengths calls for solutions beyond purely parallelization exploration (Section \ref{sec:results}).
We also study how different generations of GPUs and other commodity hardware platforms impact overall training efficiency and follow up with a future technologies scaling study by showing the effects of asymmetrically improving systems components like compute efficiency, memory capacity and bandwidth, and hierarchical interconnect bandwidth (Section \ref{sec:results}).

The main contributions of this work are as follows:

\begin{itemize}
    \item We propose a performance model that enables agile exploration of the distributed ML training and deployment design space. 
    Our performance model targets both implemented and future models alike, enabling accurate performance estimation with different model architectures, tasks, hardware devices, and distributed systems.
    \item We show model-level insights on how parallelization strategies interact with DLRM and its transformer and mixture-of-experts variants.
    We show how asymmetric compute and communication requirements from transformer and mixture-of-experts components lead to different optimal parallelization strategies.
    Additionally, we demonstrate the limits of solely optimizing parallelization strategies on LLMs of increasing context length.
    \item We show that to improve large ML model training and inference throughput, hardware specifications across compute, memory, and interconnect have to be concurrently improved.
\end{itemize}

We have open-sourced MAD-Max and sample experiment to enable follow-on work for modeling the interaction between parallelization strategies, models, tasks, and distributed systems on ML system performance.

%% file: text_CR/background.tex
\section{Background}~\label{sec:background}
In this section, we introduce a suite of model architectures across both recommender systems and LLMs.
We then outline three tasks for these models: pre-training, fine-tuning, and inference (Section \ref{ssec:bgd_models_tasks}).
Lastly, we discuss the parallelization strategies currently used to map the workloads (i.e., model and task) onto the distributed systems (Section \ref{ssec:bgd_parallelization_strategies}).

\begin{figure*}[t!]
    \centering
    \includegraphics[width=\linewidth]{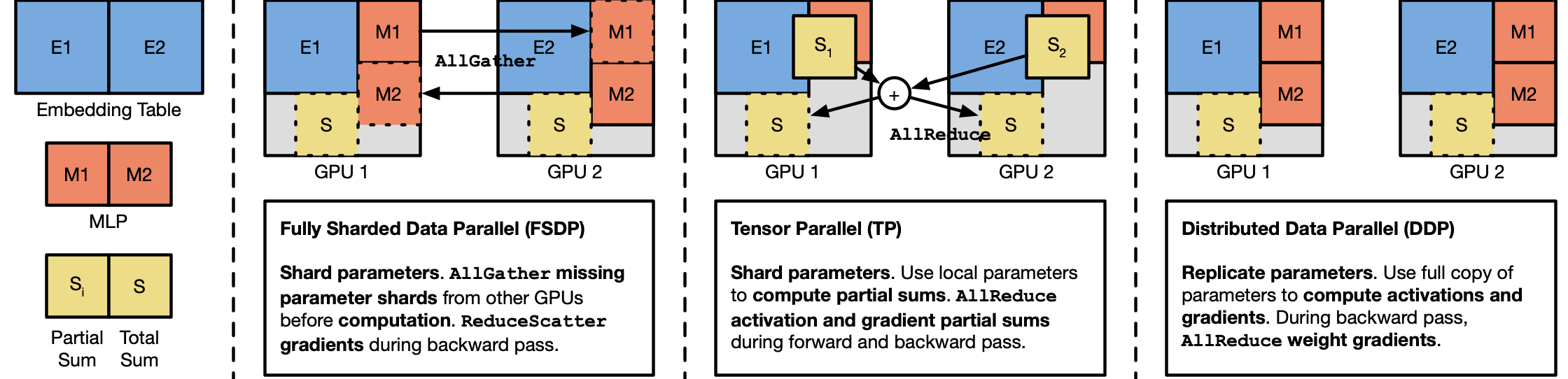}
    \caption{For recommendation models, applying FSDP, TP, or DDP on an MLP layer requires either sharding or replicating parameters and communicating either parameters (orange) or partial sums (yellow).
    In this example, the embedding table's prohibitively large capacity requires it to be sharded.}
    \label{fig:parallelization}
\end{figure*}

\subsection{Models and Tasks}~\label{ssec:bgd_models_tasks}
Deep learning based recommender systems and LLMs follow the general model architecture of representing categorical inputs as embedding vectors and then processing these embedding vectors with model-specific computation layers.
This means that there are many shared components that are emphasized to different degrees by each model: embedding tables, multilayer perceptrons (MLPs), and more intricate dense processing layers like transformer blocks.
We focus on the following five classes of models throughout the paper:
\begin{enumerate}
    \item \textbf{DLRM.} 
    The canonical at-scale recommendation model takes in dense and sparse features. 
    Dense features, such as, user age and current time, are processed by MLP layers while sparse categorical features are processed as lookups into large embedding tables.
    These results are then fed into a feature interaction layer, where these intermediate values are either concatenated or multiplied with one another via dot products~\cite{wang2017deep, wang2021dcnv2}. 
    The result of this feature interaction layer is then fed into MLP layers to generate predictions like Click-Through Rate (CTR)~\cite{naumov2019dlrm}.
    For many large-scale DLRM models, storing and communicating trillion-parameter scale embedding tables is the primary system bottleneck~\cite{gupta2020deeprecsys, gupta2020architectural, hsia2020cross, hsia2023mprec, ke2020recnmp, kwon2019tensordimm, mudigere2021zionex, sethi2022recshard, wilkening2021recssd}.
    \item \textbf{DLRM-Transformer.} 
    As sparse features for recommendation models increase in complexity, corresponding model architectures have also evolved to better model implicit relationships between sparse features.
    Some DLRM variants replace concatenation and dot-product based feature interactions with transformer encoder layers that model higher-order interactions and sequential relationship between sparse features. 
    Others use transformer-style feature interaction layers to tackle challenges like behavior sequence modeling and personalized re-ranking~\cite{chen2019alibaba, pei2019alibaba, zhang2022dhen}.
    From a systems perspective, transformer layers increase both compute and computation-communication overlap opportunities.
    \item \textbf{DLRM-MoE.}
    In the context of DLRMs, applying Mixture-of-Experts (MoE) creates parallel Top MLPs that are conditionally activated based on feature interactions~\cite{zhao2019mtwnd}.
    Because only a fraction of experts are active for each sample,
    DLRM-MoE increases model capacity and expert-to-expert communication while scaling computation at a lower rate.
    \item \textbf{LLM.}
    Large language models (LLMs) also use the \textit{``look up embeddings then process them"} architecture~\cite{brown2020gpt3, hoffmann2022chinchilla, radford2019gpt2, touvron2023llama, zhang2022opt}.
    However, instead of using user and content categorical features, LLMs convert \textit{tokens} -- character sequences -- to input embeddings.
    Subsequent processing layers use alternating self-attention and feed-forward layers~\cite{vaswani2017attention}.
    Unlike DLRMs, advancements in LLM modeling have been more focused on the processing layers than embeddings, reinforcing the importance of compute in LLM execution.
    \item \textbf{LLM-MoE.}
    In the context of LLMs, one way to apply MoE is to replace the feed-forward layer in transformer blocks with experts.
    By applying this technique, the FLOPs per token will grow at a slower rate than overall model capacity while scaling up the model, leading to enabling efficient training and inference.
    While FLOPs becomes less of a concern, the non-blocking expert-to-expert communication that can be present during training presents systems challenges.
\end{enumerate}

In terms of tasks, we are interested in pre-training, fine-tuning and inference.
Pre-training stresses all of compute, memory capacity, and communication as it involves both forward and backward passes -- along with retaining intermediate activations from the forward pass.
The requirements of fine-tuning are a subset of pre-training, as
the frozen parameters of a model do not require updates.
Inference only requires the forward pass so compute is usually proportionally larger.

\subsection{Parallelization Strategies}~\label{ssec:bgd_parallelization_strategies}
A model layer can be either replicated or sharded across devices.
We explore the following parallelization strategies (Figure \ref{fig:parallelization} illustrates forward pass execution):
\begin{enumerate}
    \item \textbf{Fully Sharded Data Parallelism (FSDP).}
    Parameters are \textit{sharded} across devices.
    Before forward and backward pass, missing parameter shards are gathered via \texttt{AllGather}.
    During backward pass, weight gradients are reduced and sharded via \texttt{ReduceScatter}.
    \item \textbf{Tensor Parallelism (TP).}
    Parameters are \textit{sharded} across devices.
    During forward pass, each device uses its parameter shard to compute partial sums that are then aggregated via \texttt{AllReduce}.
    Same principle is applied for backward pass for gradients.
    \item \textbf{Distributed Data Parallelism (DDP).}
    Parameters are \textit{replicated} across devices.
    During forward pass, each device acts independently for computation.
    During backward pass, devices \texttt{AllReduce} weight gradients.
\end{enumerate}
We apply one parallelization strategy for each layer type.
Figure \ref{fig:parallelization} depicts applying different parallelization strategies on an MLP layer and vanilla \textbf{model parallel (MP)} sharding for the embedding tables.
Additionally, parallelization strategies can be applied hierarchically for multi-node systems, creating $N$-D parallelism strategies.

%% file: text_CR/characterization.tex
\begin{figure*}[t!]
    \centering
    \includegraphics[width=\linewidth]{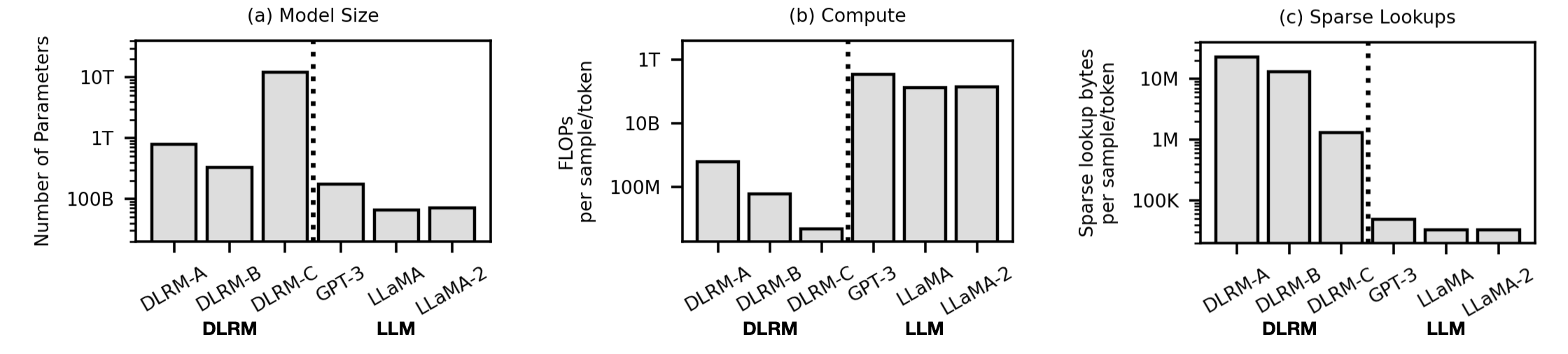}
    \caption{For large ML models, the requirements for key system resources -- \textbf{(a)} capacity, \textbf{(b)} compute, \textbf{(c)} bandwidth -- vary by orders of magnitude.}
    \label{fig:model_char}
\end{figure*}

\begin{figure*}[t!]
    \centering
    \includegraphics[width=\linewidth]{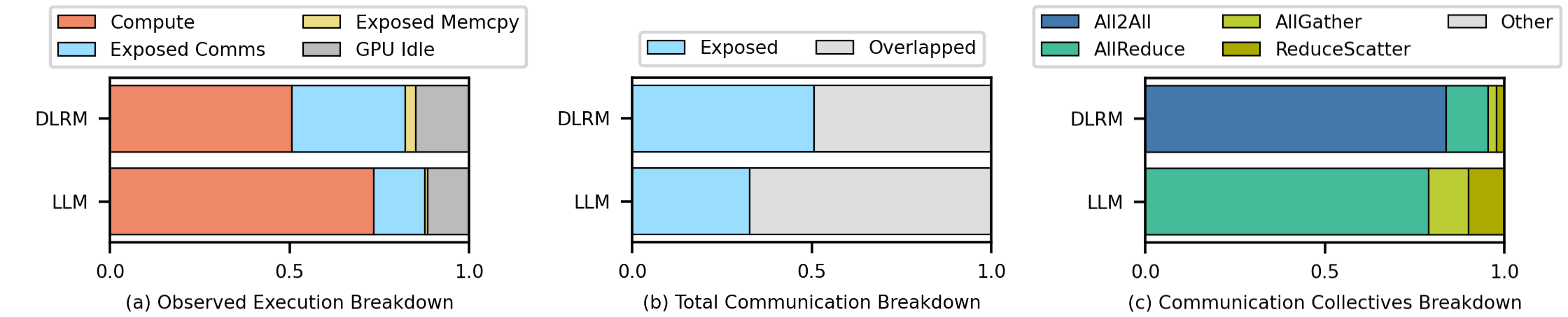}
    \caption{\textbf{(a)} Compute and exposed communication make up the majority of observed at-scale training cycles.
    \textbf{(b)} The degree of communication overlapped with computation and data loading is workload dependent.
    Higher degree of overlap indicates better latency hiding of communication collectives.
    \textbf{(c)} Breakdown of communication collectives also varies by workload.}
    \label{fig:comms_characterization}
\end{figure*}

\section{Characterization}~\label{sec:characterization}
In this section, we first characterize a suite of real-world large ML models with respect to model capacity, parameter breakdowns, FLOPs, and memory bandwidth characteristics (Section \ref{ssec:chr_models}).
To get a better understanding of the models' communication requirements, we conduct a fleet-wide characterization of at-scale training experiments (Section \ref{ssec:chr_comms}).

\subsection{Individual Model Characterization}~\label{ssec:chr_models}
We first quantify the difference in compute, memory capacity, and bandwidth requirements between six real-world recommendation models and LLMs: DLRM-\{A, B, C\}, GPT-3 175B, LLaMA-65B, LLaMA 2-70B.
Figure \ref{fig:model_char} quantifies this diversity of requirements with two key observations:

\textbf{\underline{O1:}
Parameter count -- and allocation across model layers -- varies by orders of magnitude between models, impacting system capacity requirements}.
Recommendation models contain significantly more parameters than LLMs (Figure \ref{fig:model_char} (a)).
Despite variation in parameter count across LLMs, GPT-3 consists of roughly 2--68$\times$ fewer parameters as compared to recommendation models.
Training and deploying these recommendation models and LLMs require multi-node distributed systems, yet the size of the target model governs how many devices (i.e., GPUs) are required to fit the entire model and the viable set of scale-out parallelization strategies.

Additionally, virtually 100\% of parameters in recommendation models are used for embeddings while almost all LLM parameters are dedicated to compute. 
This reflects the transformer-heavy computation of current LLM architectures, in contrast to embedding-driven, recommendation model architectures for at-scale personalization. 

\textbf{\underline{O2:}
Recommendation models require fewer FLOPs per sample as compared to LLMs, yet require \textgreater20$\times$ higher memory bandwidth for sparse lookups.}
Figures~\ref{fig:model_char} (b, c) illustrate how recommendation models and LLMs show opposite trends for compute requirements and sparse lookup bandwidth.
Sparse lookup bandwidth requirements for recommendation models far surpass LLMs -- a fact that is consistent with how recommendation models have a higher proportion of parameters dedicated to embeddings.
However, the opposite is true for compute requirements, as LLMs require significantly higher FLOPs per sample.
As discussed in Section \ref{sec:design}, these varying system requirements play an important role in the design of an optimal parallelization strategies.

\begin{figure*}[t!]
    \centering
    \includegraphics[width=\linewidth]{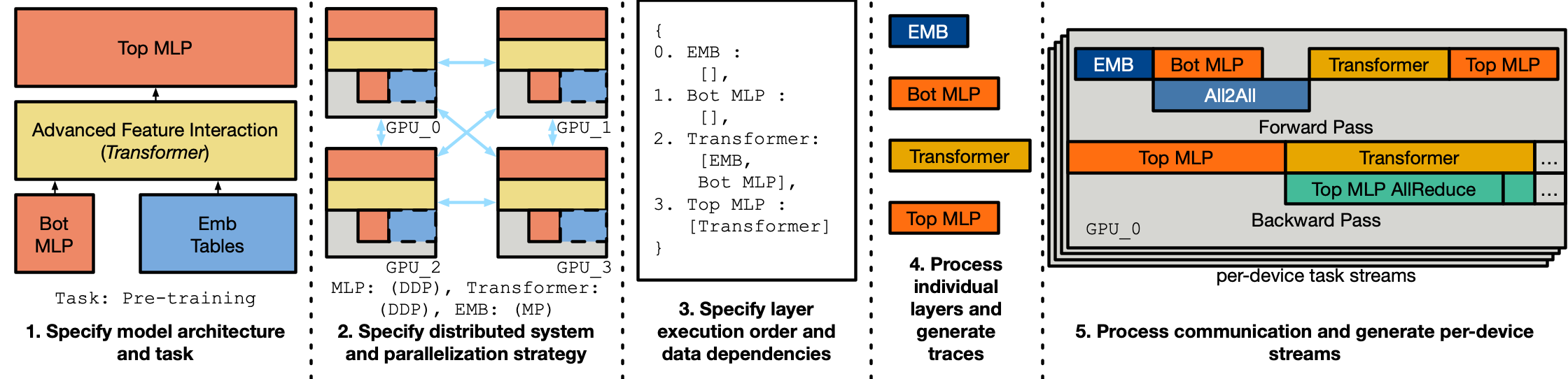}
    \caption{Our performance model works in five stages.
    After workload specifications and layer execution orders are established, traces for individual layer execution are generated and then combined with required communication collectives to form complete computation and communication streams.}
    \label{fig:design_flow}
\end{figure*}

\subsection{Fleet-wide Communication Characterization}~\label{ssec:chr_comms}
In addition to model-level characterization, we look at fleet-wide model training.
We observe, over an extended period of time, the importance of communication for training the latest DLRM-style models and LLMs.
Figure \ref{fig:comms_characterization} quantifies the role of communication with two key observations: 

\textbf{\underline{O3:}
Compute and exposed communication make up the majority of observable training GPU cycles.}
Compute, defined as cycles with either device computation or memory lookups (orange) and exposed communication, defined as cycles with only inter-device communication (blue), make up \textgreater82\% of all observable training GPU cycles for both DLRM and LLMs (Figure \ref{fig:comms_characterization} (a)).
The rest of the cycles are attributed to host-device communication -- exposed memcpy (yellow) -- and inactivity due to data ingestion, kernel launch overhead, etc. -- GPU idle (grey).
\textit{From this observation, we focus our performance modeling efforts on predicting the expected behavior of compute and communication cycles.}

\textbf{\underline{O4:}
Differences in model architectures and parallelization strategies impact both the amount of compute-communication overlap and the types of communication collectives used.}
When model training spans multiple devices, replicating or sharding model components leads to communication calls involving parameters, activations and/or gradients.
Being able to overlap these communication calls with computation so that the hardware devices are doing useful work is important for utilization.
Figure~\ref{fig:comms_characterization} (b) shows that $\sim$50\% of communication calls for DLRM training are overlapped with computation, whereas \textgreater65\% of communication calls for compute-dominated LLMs are overlapped.

Figure \ref{fig:comms_characterization} (c) shows the spread of different communication collectives during training.
For DLRM models, \texttt{All2All} is heavily emphasized while LLMs spend the majority of their communication cycles on \texttt{AllReduce}.
This is a direct result of model architecture difference, and thus active parallelization strategy.
Since DLRMs require large amounts of sparse lookups from sharded embedding tables, the per-device unique embedding lookups have to be distributed to each device via \texttt{All2All}.
On the contrary, LLMs have fewer parameters and are more amenable to replication of compute parameters, allowing for DDP opportunities that require \texttt{AllReduce} for aggregating weight gradients.

In this section, we characterize real-world large ML models from model architecture and distributed training perspectives.
From Section \ref{ssec:chr_comms} we see that model architectures and the way in we map them onto distributed systems significantly impacts system resource utilization, and thus overall performance.
To better understand how to best map current and future large ML models onto different distributed systems, we propose an agile, at-scale accurate performance model.

%% file: text_CR/design.tex
\section{Proposed Design}~\label{sec:design}
\rewritedesign{
In this section, we outline the structure of our performance model -- MAD-Max -- for simulating distributed ML workloads.
We begin with an overview of the model's design and the key assumptions it relies on, highlighting the role of execution traces in modeling the iterative behavior of large-scale ML tasks (Section~\ref{ssec:design_overview}).
Then, we discuss how the model processes individual ML model layers according to their key characteristics (Section~\ref{ssec:design_layers}).
We conclude by explaining the integration of these individually-processed layers into a unified computation and communication model that addresses the communication requirements dictated by the chosen parallelization strategy (Section~\ref{ssec:design_streams}).}

\subsection{\rewritedesign{Design Overview}}~\label{ssec:design_overview}
\rewritedesign{
Our performance model, illustrated through a DLRM-Transformer case in Figure~\ref{fig:design_flow}, is predicated on the notion that ML model layers, when treated as discrete blocks, can be used to create \textbf{per-device execution traces} for emulating the per-iteration behavior of distributed ML workloads.
An ``\textbf{execution trace}" in this context refers to a detailed record capturing the sequence and duration of both compute and communication events (i.e., streams) on each device.
To simulate the per-iteration behavior of a distributed ML workload, MAD-Max constructs a dependency graph of layers, generates per-layer compute traces, and then pieces together the compute traces with traces of parallelization strategy specific communication collectives to form complete compute and communication streams.
From per-iteration behavior, the performance model estimates overall throughput and other end-to-end serialized and overlapped execution breakdowns.}

Users have to provide JSON files for: 1) model architecture via layer-specific configurations (e.g., number of MLP layers, embedding table dimension, number of transformer layers and heads), 2) distributed system specifications (e.g., Tensor Float (TF32) utilization, HBM peak bandwidth, \texttt{AllReduce} intra-node interconnect utilization), and 3) task and parallelization strategy (e.g., pre-training/fine-tuning/inference, intra-/inter-node parallelization strategies).

With these configurations, individual layers are first processed by their primary system requirements.
Examples include estimating embedding bag execution by the amount of embeddings to look up and per-GPU high-bandwidth memory (HBM) memory bandwidth and the time it takes to execute a transformer encoder layer by TF32 compute throughput. 
Based on the replication and sharding specified by the target parallelization strategy, the required communication collectives are processed by collective-specific intra- (e.g., NVLink) and inter- (e.g., Infiniband, RDMA over Converged Ethernet (RoCE)) node communication bandwidths.

We take into account task-level requirements (i.e., pre-training/fine-tuning/inference) to construct per-device computation and communication streams with data dependencies and potential computation-communication overlap.

\textbf{Assumptions:}
\begin{itemize}
    \item Since we focus on large-models, target distributed systems are multi-device in nature. 
    For multi-device execution, a first-order analysis of execution behavior and overall performance can be estimated via modeling per-node layer execution and inter-node parallelization communication.
    Kernel-level improvements (e.g.,~\cite{nvidia2023transformerengine}), while not the focus of this work, can be effectively modeled as increased compute and memory lookup utilization.
    \item The performance model assumes that the entire model can be fit onto the training/inference devices (i.e., when sharded, the model can fit onto GPUs).
    Recent high-performance training platforms target this design point~\cite{mudigere2021zionex}.
    Design points where model parameters have to be shuffled back and forth between CPU and device are currently unsupported.
    \item Device-host communication (e.g., CPU-GPU data loading) is relatively a second-order consideration and mostly overlapped and hidden between training/inference iterations.
    This observation is shared in~\cite{mudigere2021zionex} and our fleet-wide characterization in Section \ref{ssec:chr_comms}, Figure \ref{fig:comms_characterization}.
\end{itemize}

\subsection{Processing Individual Model Layers}~\label{ssec:design_layers}
Layers are processed by their main system requirement. For example, we illustrate how MLP and embedding bag performance are estimated differently in Figure~\ref{fig:design_flow}.  

\textbf{Compute Blocks.}
Assuming that compute time is the main bottleneck for MLPs, we estimate compute time per layer as:
\begin{center}
    $\sim$ (FLOPs per layer) / [(GPU peak FLOPS) * Compute utilization]
\end{center}
where FLOPs per layer is determined by the MLP layer's dimensions and target batch size.
GPU peak FLOPS are heavily dependent on data type (e.g., 32-bit, 16-bit FP/TF/BF) and whether or not tensor cores are enabled.
We incorporate compute utilization -- or in the case of GPUs, SM utilization/occupancy -- as a factor in [0,1]. 
Typical compute utilization factors for A100s on layers in our models of interest are $\sim$70\%.
We adopt a similar approach for modeling self-attention and fully-connected (FC) layers found in transformer layers, where FLOPs per layer is estimated by additional factors such as attention dimension and context length.

\textbf{Embedding Bags.}
Assuming that lookup time is the main bottleneck for embedding bags, we estimate lookup time as:
\begin{center}
    $\sim$ (Lookup bytes per GPU) / [(HBM BW) * HBM utilization]
\end{center}
where Lookup bytes is determined by the number of embedding tables, number of lookups per embedding table, embedding dimension, and embedding precision.
Lookup bytes per GPU is highly parallelization strategy dependent.
In this case, we assume that the embedding table is evenly sharded across GPUs in terms of both capacity and number of lookups. 
If the number of lookups are unevenly distributed between GPUs, we can adjust the lookup bytes per GPU on a per-GPU basis~\cite{sethi2022recshard}.
HBM utilization is a factor between [0,1] and typical values for embedding bags of interest are $\sim$80\% for A100s.

\subsection{Piecing Together Computation and Comm. Streams}~\label{ssec:design_streams}
\textbf{Specifying Explicit Execution Order.}
To generate per-device traces for different ML tasks, an explicit execution priority must be established for the different layers.
In Figure~\ref{fig:design_flow}, we can establish the order as such (1) Embedding, (2) Bottom MLP, (3) Transformer, (4) Top MLP.
During backward pass, the execution order will be reversed.
If the target task is fine-tuning, we also specify frozen layers, reducing unnecessary computation and communication of certain weight gradients.

\begin{figure}[t!]
    \centering
    \includegraphics[width=\linewidth]{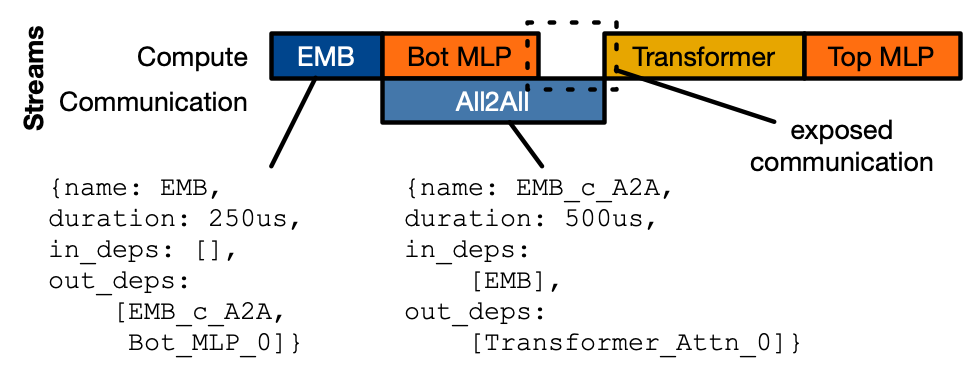}
    \caption{Sample generated GPU compute and communication streams with labeled exposed communication.}
    \label{fig:sample_streams}
\end{figure}

\textbf{Generating Parallelization-Specific Streams.}
An explicit execution order by itself is not enough to construct accurate streams.
A target parallelization strategy is required to specify the required communication collectives.
Explicit data dependencies, along with parallelization strategy determine the blocking/non-blocking nature of the communication calls.
In Figure \ref{fig:design_flow}, MLP and transformer layers are distributed via DDP while embedding tables are distributed via sharding.

Figure \ref{fig:sample_streams} illustrates generated forward pass streams from our DLRM-Transformer example.
We see that the traces are slotted into a compute stream and communication stream.
Each trace will have dependencies that come explicitly from execution annotations and implicitly from underlying parallelization strategies.
For example, \texttt{EMB} has an explicit output dependency of \texttt{Bot\_MLP\_0} and implicit output dependency of \texttt{EMB\_c\_A2A} from sharding the embedding table. 
\texttt{EMB\_c\_A2A} is blocking since \texttt{Transformer\_Attn\_0} needs \texttt{EMB\_c\_A2A}’s results.

\textbf{Estimating Communication Collective Execution.}
We estimate All2All execution as:
\begin{center}
    $\sim$ (“SendCount” Bytes per GPU) / (Effective All2All BW)
\end{center}
where “SendCount” Bytes per GPU is the average number of bytes sent by each GPU to every other GPU.
“SendCount” Bytes per GPU is dependent on not only ``Lookup bytes per GPU" but also the sharding degree and number of devices.
\rewritedesign{
Since the All2All NCCL implementation is composed of individual point-to-point \texttt{Send()} and \texttt{Recv()} calls, it is bound by the slowest level of interconnect~\cite{woolleynccl}.}
Thus, for baseline DGX systems, Effective All2All BW is set as that of either Infiniband or RoCE (i.e., whatever interconnect fabric is used to connect nodes of GPUs).
For other cases, like an 8-GPU system, Effective All2All BW may be NVLink BW.

Likewise, we can generate a similar set of traces for the backward pass.
Since the MLP and transformer layers are parallelized via DDP, we have non-blocking AllReduce communication calls during the backward pass.
The AllReduce calls are for aggregating per-layer weight gradients and are thus non-blocking (i.e., they are not on the critical path for backpropagation).
We estimate the non-blocking AllReduce calls for weight gradient calls as:
\begin{center}
    $\sim$ (“SendBuffer” Bytes / GPU) / (Effective AllReduce BW)
\end{center}
where “SendBuffer'' Bytes is the total number of bytes sent by each GPU and is directly proportional to the number of parameters in each layer.
\rewritedesign{
Effective AllReduce BW is a ratio of intra-node communication (e.g., NVLink) bandwidth and inter-node communication (e.g., Infiniband or RoCE) bandwidth since data is communicated on both classes of channels for the NCCL implementation~\cite{woolleynccl}.}
The exact ratio between the two communication technologies is dependent on factors like the number of nodes and NCCL implementation version (e.g., ring vs. tree). 
We use real hardware measurement data via to understand what these effective interconnect ratios and bandwidths are in practice.
Large-scale training also often exhibits non-constant bandwidth across intra- and inter-node hierarchies.
We also consider AllGather and ReduceScatter communication calls, which are required in FSDP and TP.

\textbf{Computation-Communication Overlap.}
We maintain separate compute and communication streams and overlap traces with no data dependencies.
We also assume GPU kernels are launched whenever data dependencies are resolved.
Ideally, we want to maximize compute-communication overlap.
However, as demonstrated in Figure~\ref{fig:sample_streams}, there is a segment of exposed communication for the All2All operation, indicating the GPU's compute and memory resources are idle and underutilized.

MAD-Max allows us to both identify combinations of kernels and parallelization strategies that lead to exposed communication and experiment with different parallelization strategies to decrease exposed communication segments.
Optimizing for computation-communication overlap is an important objective across multi-node, large-scale ML workloads.
Currently, 14$\sim$32\% of GPU cycles on the training clusters come from exposed communication (Figure \ref{fig:comms_characterization}).

%% file: text_CR/methodology.tex
\input{tables/method_validation}

\begin{figure}[t!]
    \centering
    \includegraphics[width=\linewidth]{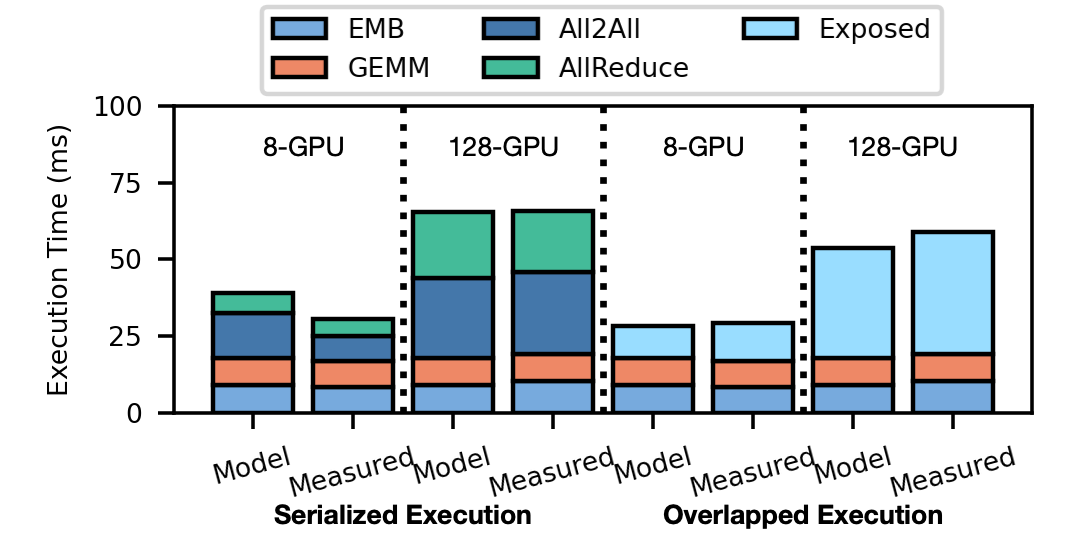}
    \caption{DLRM-A serialized and overlapped execution validation for 8-, 128-GPU training.}
    \label{fig:dlrma_validation}
\end{figure}

\section{Experimental Methodology}~\label{sec:methodology}
This section describes our validation and outlines the design space of this study, including variations of real-world models, hierarchical parallelization strategies, and hardware platforms.

\begin{figure}[t!]
    \centering
    \includegraphics[width=\linewidth]{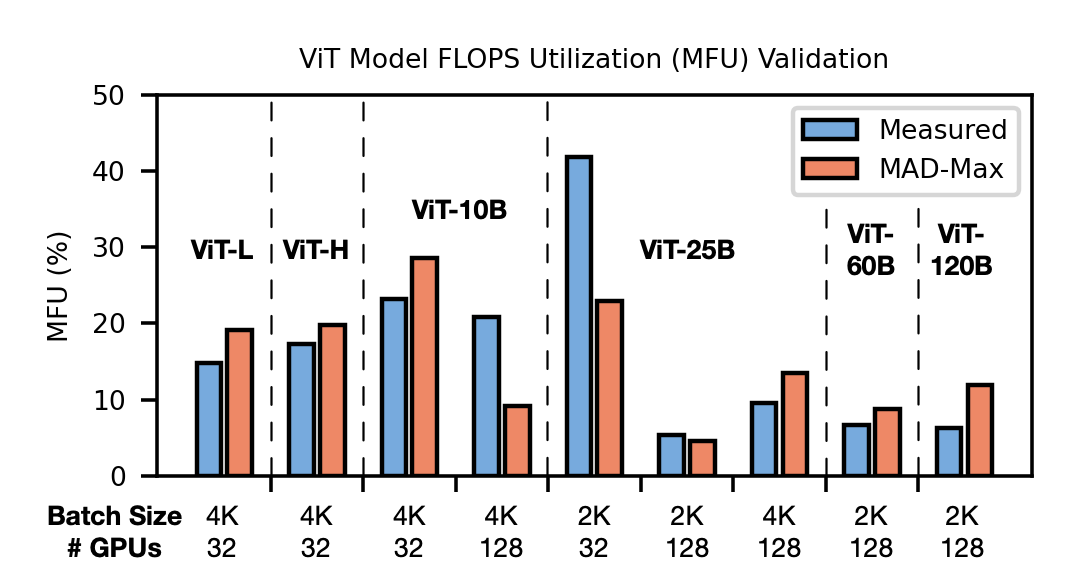}
    \caption{\srt{ViT validation across different model sizes, global batch sizes, and number of GPUs on AWS \texttt{p4d\_24xlarge} instances.}}
    \label{fig:vit_validation}
\end{figure}

\begin{figure}[t!]
    \centering
    \includegraphics[width=\linewidth]{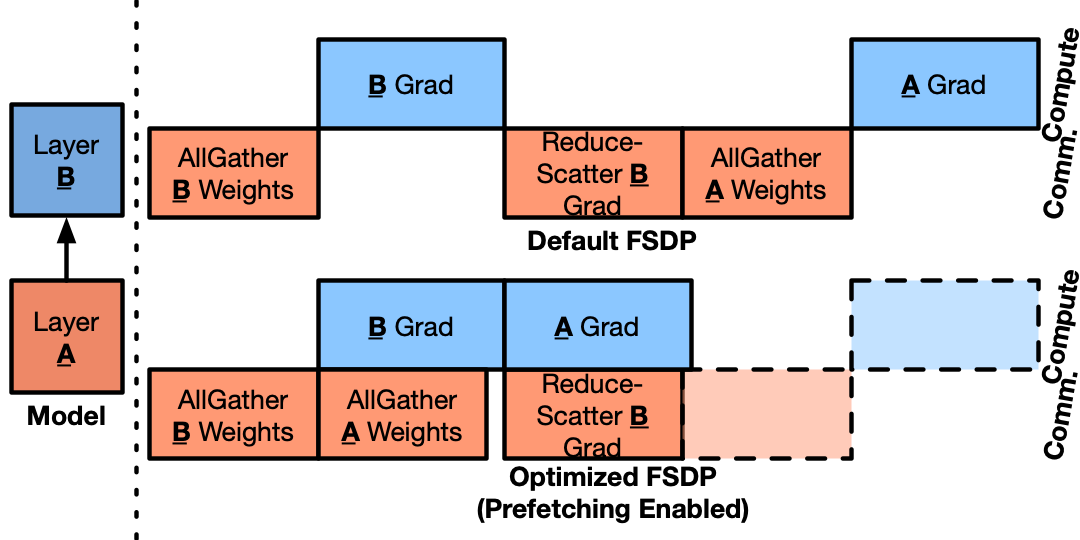}
    \caption{\srt{Optimized FSDP implementation with prefetching that we validate against for production LLaMA training traces.}}
    \label{fig:optimal_fsdp}
\end{figure}

\input{tables/method_models}
\input{tables/method_hw}

\textbf{Performance Model Validation.}
Table \ref{tbl:method_validation} lists validation points of various first-order execution metrics across real, measured recommendation and LLM training experiments.
For DLRM-A training~\cite{mudigere2021zionex}, we validate the performance model over the key dimensions of serialized iteration time, \% communication exposed, and training throughput for 96.89, 91,62, and 99.17\% modeling accuracy, respectively.
Figure \ref{fig:dlrma_validation} compares the execution time of DLRM-A training in detail across 8- and 128-A100 ZionEX platforms.
We validate serialized execution to check layer execution and collectives' volumes, overlapped execution to check at-scale latency-hiding opportunities and systems of different number of nodes to observe networking scaling effects.
For DLRM-B training, our model reports 3.05 MQPS whereas the measured throughput is 3.4 MQPS for 89.7\% modeling accuracy.

For the largest LLaMA configuration (LLaMA-70B), our performance model estimates training time for all 1.4T tokens to take 19.21 days as opposed to the reported 21 days in \cite{touvron2023llama}.
For this use-case, we use the same hardware platform as reported in~\cite{touvron2023llama} (i.e., 2048 80GB HBM A100s).
We also validate the aggregate GPU Hours to train for 306k steps, resulting in 84.66\% modeling accuracy.

\srt{Figure~\ref{fig:vit_validation} presents additional validation points on Vision Transformer (ViT) model training across a range of model configurations, global batch sizes, and number of GPUs.
ViT models range from 300M (ViT-L) to 120B (ViT-120B) parameters and global batch size is set at either 2 or 4K for target model accuracies.
All experiments are done on AWS \texttt{p4d\_24xlarge} instances and using the baseline FSDP parallelization strategy.
We model SM utilization as a function of GPU local batch size and model layer FLOPs requirements.
Across all the data points, we get an average of 93.88\%  and median of 95.74\% accuracy for model flops utilization (MFU).}

\srt{
Figure~\ref{fig:optimal_fsdp} shows a visualization of communication and computation streams for an optimized implementation of FSDP with prefetching enabled.
In this optimized variation of FSDP, earlier layer (i.e., Layer A) weight AllGathers are prefetched and overlapped with later layer (i.e., Layer B) gradient computation, leading to overall execution time speedup.
We validate this collective-level optimization in MAD-Max against a production implementation and corresponding GPU traces.
For a specific LLaMA pre-training run using this optimization, we observe 98\% communication overlap against a predicted 93\% communication overlap for MAD-Max simulation.}

\begin{figure*}[t!]
    \centering
    \includegraphics[width=\linewidth]{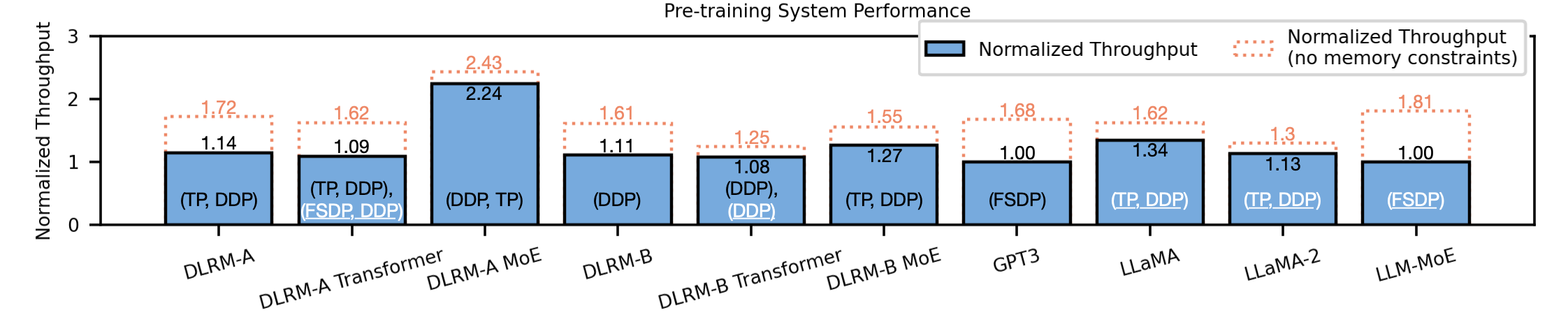}
    \caption{We can improve pre-training performance over FSDP baseline by applying intra- and inter-node parallelization strategies for base dense and transformer layers separately.
    Throughput-optimal parallelization strategies are listed in (intra-, inter-) order.
    Black and white, underlined text refer to recommendation base dense and transformer layers, respectively.}
    \label{fig:overall_results}
\end{figure*}

\textbf{Model Variations.}
Table \ref{tbl:method_models} lists the suite of large ML models explored in Section \ref{sec:results}.
We explore transformer and MoE variants of real-world DLRM-A and DLRM-B.
The transformer feature interaction variants have 4 layers and a down-sampled sequence length of 80.
MoE variants are configured with 16 experts (2 active) per layer.
For the LLM models, we follow specifications in~\cite{brown2020gpt3, touvron2023llama, touvron2023llama2}.
For LLM-MoE, we explore a hypothetical 1.8T parameter model with 16-(2 active)way MoE for the MLPs in transformer blocks.
We use fixed global batch sizes as specified in prior studies~\cite{mudigere2021zionex, touvron2023llama} to maintain target model accuracy.

\textbf{Design Space Exploration.}
We use FSDP~\cite{zhao2023fsdp} as the baseline due to its wide adoption and ability to best guarantee training feasibility by minimizing memory footprint.
We explore valid hierarchical parallelism strategies at intra- and inter-node levels, considering combinations of DDP, FSDP, and TP.
For hardware, unless otherwise stated, we use training systems from prior case studies~\cite{mudigere2021zionex, touvron2023llama} (Table \ref{tbl:method_hw}).
We also explore implications of using H100 and H100 SuperPOD systems by additional simulations replacing our A100-based models with H100 specifications~\cite{nvidia2023h100, nvidia2023hgxh100} -- Table~\ref{tbl:other_hw}.

\srt{\textbf{Validation Efforts.}
The efforts behind large-scale GPU validation experiments can easily add up.
Our DLRM-A and -B validation experiments that were critical for tuning \textit{MAD-Max} (Table~\ref{tbl:method_validation}, Figure~\ref{fig:dlrma_validation}) took $\sim$64K aggregate A100 GPU hours.
Running the same experiments on AWS \texttt{p4d\_24xlarge} EC2 instances -- which also have 4$\times$ lower inter-node interconnect bandwidth compared to systems enumerated in Table~\ref{tbl:method_hw} -- would amount to even more aggregate GPU hours.
Additionally, the LLaMA and ViT validation experiments (Table~\ref{tbl:method_validation}, Figure~\ref{fig:vit_validation}), which were run across a range of 32 to 2048 GPUs, would require comparable aggregate GPU hours.
}

%% file: tables/method_validation.tex
\begin{table}[t!]
\vspace{5mm}
\begin{center}\resizebox{\linewidth}{!}{
\begin{tabular}{|c||c|c|c|c|}
\hline
\textbf{}                           & \textbf{Evaluation Metric}                                                      & \textbf{\begin{tabular}[c]{@{}c@{}}Measured\\ Result\end{tabular}} & \textbf{\begin{tabular}[c]{@{}c@{}}Performance\\ Model \\ Result\end{tabular}} & \textbf{\begin{tabular}[c]{@{}c@{}}Modeling \\ Accuracy \\ (\%)\end{tabular}} \\ \hline
\multirow{3}{*}{\textbf{DLRM-A}}    & Serialized Iteration Time (ms)                                                  & 67.40 ms                                                           & 65.30 ms                                                                       & 96.89\%                                                                       \\ \cline{2-5} 
                                    & \% Communication Exposed (\%)                                                   & 82.37\%                                                            & 75.46\%                                                                        & 91.62\%                                                                       \\ \cline{2-5} 
                                    & \multirow{2}{*}{Throughput (MQPS)}                                              & 1.2 MQPS~\cite{mudigere2021zionex}                                                           & 1.21 MQPS                                                                      & 99.17\%                                                                       \\ \cline{1-1} \cline{3-5} 
\textbf{DLRM-B}                     &                                                                                 & 3.4 MQPS~\cite{mudigere2021zionex}                                                           & 3.06 MQPS                                                                      & 90\%                                                                          \\ \hline
\multirow{2}{*}{\textbf{LLaMA-70B}} & \begin{tabular}[c]{@{}c@{}}GPU Hours for 306k steps\\ (2048 A100s)\end{tabular} & \begin{tabular}[c]{@{}c@{}}1,022,361\\ Hrs\end{tabular}            & \begin{tabular}[c]{@{}c@{}}863,397\\ Hrs\end{tabular}                          & 84.66\%                                                                       \\ \cline{2-5} 
                                    & Days to Train 1.4T Tokens                                                       & 20.83 Days~\cite{touvron2023llama}                                                         & 19.21 Days                                                                     & 92.27\%                                                                       \\ \hline
\end{tabular}}
\end{center}
\caption{Validation of various first-order execution metrics.}
\label{tbl:method_validation}
\end{table}

%% file: tables/method_models.tex
\begin{table*}[t!]
\vspace{2.5mm}
\begin{center}\resizebox{\linewidth}{!}{
\begin{tabular}{|c||cccccc|cccc|}
\hline
\textbf{}                                                                               & \multicolumn{1}{c|}{\textbf{DLRM-A}~\cite{mudigere2021zionex}} & \multicolumn{1}{c|}{\textbf{DLRM-A Transformer}} & \multicolumn{1}{c|}{\textbf{DLRM-A MoE}} & \multicolumn{1}{c|}{\textbf{DLRM-B}~\cite{mudigere2021zionex}} & \multicolumn{1}{c|}{\textbf{DLRM-B Transformer}} & \textbf{DLRM-B MoE} & \multicolumn{1}{c|}{\textbf{GPT-3}~\cite{brown2020gpt3}} & \multicolumn{1}{c|}{\textbf{LLaMA}~\cite{touvron2023llama}} & \multicolumn{1}{c|}{\textbf{LLaMA2}~\cite{touvron2023llama2}} & \textbf{LLM-MoE} \\ \hline
\textbf{\# Parameters}                                                                  & \multicolumn{2}{c|}{793B}                                                               & \multicolumn{1}{c|}{795B}                & \multicolumn{2}{c|}{332B}                                                               & 333B                & \multicolumn{1}{c|}{175B}           & \multicolumn{1}{c|}{65.2B}          & \multicolumn{1}{c|}{70B}             & 1.8T             \\ \hline
\textbf{\begin{tabular}[c]{@{}c@{}}FLOPs\\ per sample/token\end{tabular}}               & \multicolumn{1}{c|}{638M}            & \multicolumn{1}{c|}{2.6B}                        & \multicolumn{1}{c|}{957M}                & \multicolumn{1}{c|}{60M}             & \multicolumn{1}{c|}{2.1B}                        & 90M                 & \multicolumn{1}{c|}{350B}           & \multicolumn{1}{c|}{130.4B}         & \multicolumn{1}{c|}{140B}            & 550B             \\ \hline
\textbf{\begin{tabular}[c]{@{}c@{}}Sparse Lookup Bytes\\ per sample/token\end{tabular}} & \multicolumn{3}{c|}{22.61 MB}                                                                                                      & \multicolumn{3}{c|}{13.19 MB}                                                                                 & \multicolumn{1}{c|}{49.2 KB}        & \multicolumn{2}{c|}{32.8 KB}                                               & 42.8 KB          \\ \hline
\textbf{Global Batch Size}                                                              & \multicolumn{3}{c|}{64K}                                                                                                           & \multicolumn{3}{c|}{256K}                                                                                     & \multicolumn{4}{c|}{2K (4M tokens)}                                                                                                 \\ \hline
\textbf{Context Length}                                                                 & \multicolumn{1}{c|}{N/A}             & \multicolumn{1}{c|}{80}                          & \multicolumn{2}{c|}{N/A}                                                        & \multicolumn{1}{c|}{80}                          & N/A                 & \multicolumn{2}{c|}{2048}                                                 & \multicolumn{1}{c|}{4096}            & 8192             \\ \hline
\end{tabular}}
\end{center}
\caption{\srt{Target recommendation models, LLMs, and their variants by key model-level characteristics.}}
\label{tbl:method_models}
\vspace{-5mm}
\end{table*}

%% file: tables/method_hw.tex
\begin{table}[t!]
\begin{center}\resizebox{\linewidth}{!}{
\begin{tabular}{|c|cc|}
\hline
                                                                                                      & \multicolumn{1}{c|}{\textbf{\begin{tabular}[c]{@{}c@{}}DLRM \\ Training System~\cite{mudigere2021zionex}\end{tabular}}} & \textbf{\begin{tabular}[c]{@{}c@{}}LLM \\ Training System~\cite{touvron2023llama}\end{tabular}} \\ \hline
\textbf{Base device}                                                                                  & \multicolumn{1}{c|}{NVIDIA A100 40GB}                                                         & NVIDIA A100 80GB                                                        \\ \hline
\textbf{Devices per node}                                                                             & \multicolumn{2}{c|}{8}                                                                                                                                                  \\ \hline
\textbf{\# nodes}                                                                                     & \multicolumn{1}{c|}{16}                                                                       & 256                                                                     \\ \hline
\textbf{Peak TF32 throughput}                                                                         & \multicolumn{1}{c|}{20 PFLOPS}                                                                & 319 PFLOPS                                                              \\ \hline
\textbf{HBM capacity}                                                                                 & \multicolumn{1}{c|}{5 TB}                                                                     & 164 TB                                                                  \\ \hline
\textbf{HBM bandwidth}                                                                                & \multicolumn{1}{c|}{199 TB/s}                                                                 & 3.96 PB/s                                                               \\ \hline
\textbf{\begin{tabular}[c]{@{}c@{}}Intra-node interconnect\\ bandwidth (unidirectional)\end{tabular}} & \multicolumn{1}{c|}{38.4 TB/s}                                                                & 614.4 TB/s                                                              \\ \hline
\textbf{Inter-node interconnect fabric}                                                               & \multicolumn{1}{c|}{RoCE}                                                                     & Infiniband                                                              \\ \hline
\textbf{\begin{tabular}[c]{@{}c@{}}Inter-node interconnect\\ bandwidth (unidirectional)\end{tabular}} & \multicolumn{1}{c|}{25.6 Tbps}                                                                & 409.6 Tbps                                                              \\ \hline
\end{tabular}}
\end{center}
\caption{\srt{Baseline distributed systems used in evaluation.}}
\label{tbl:method_hw}
\end{table}

%% file: text_CR/results.tex
\section{Evaluation Results and Analysis}~\label{sec:results}
When parallelization strategies are tailored to specific deep learning models and tasks at hand, we can achieve 8$\sim$124\% throughput improvement.
Figure~\ref{fig:overall_results} overviews pre-training throughput of key large ML models (Table~\ref{tbl:method_models}) normalized to the baseline.
We achieve, on average 65.9\% pre-training throughput improvement (blue bars) over FSDP by tuning parallelization strategies at the layer-type granularity.
The strategy that achieves optimal training throughput is indicated in parenthesis.
For example, when considering the base dense layers of DLRM-A, applying Tensor Parallelism within a node of 8 GPUs and Distributed Data Parallelism across nodes of GPUs (i.e., (\texttt{TP, DDP})) leads to optimal pre-training throughput.
In cases like DLRM-A Transformer, where both base dense and transformer layers are present, the optimal way to parallelize each type of layer may differ. 

Additionally, the orange dotted bars represent potential throughput improvements from optimizing parallelization strategies without memory constraints of current distributed systems.
The optimal parallelization strategy and its expected benefits are influenced by several factors, including the model architecture, distributed system, and task.
We highlight 10 key observations and discuss the underlying insights:

\begin{figure}[t!]
    \centering
    \includegraphics[width=\linewidth]{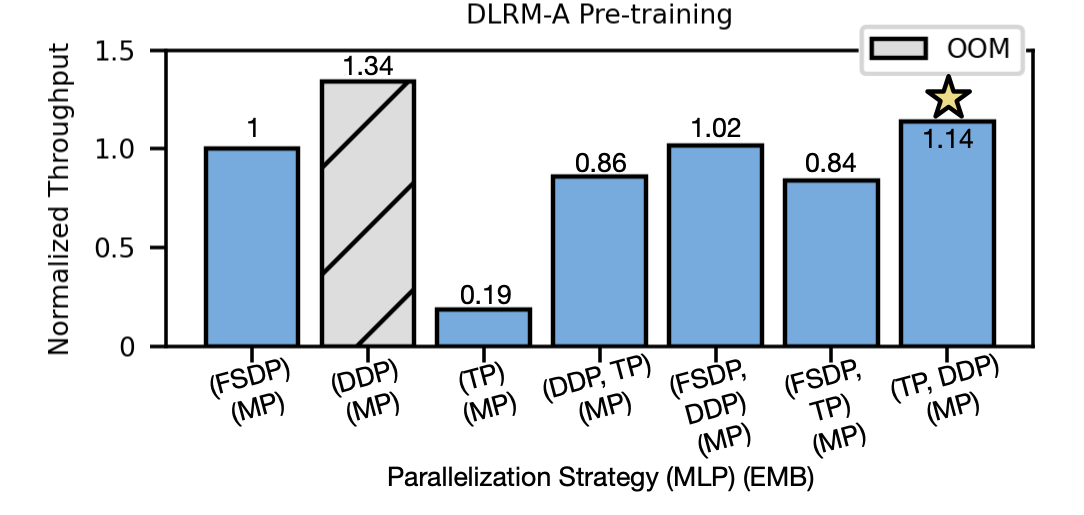}
    \caption{\textbf{DLRM-A Pre-training.}
    Considering memory capacity constraints, applying TP and DDP for intra- and inter-node parallelism, respectively on base dense layers achieves highest throughput. 
    Gray bar indicates invalid parallelism strategy due to OOM.}
    \label{fig:dlrma_pretraining}
\end{figure}

\textbf{\underline{Insight 1: [DLRM]}
Trillion-parameter embedding tables in DLRMs limit parallelization strategies for the tables to sharding, shifting overall parallelization strategy exploration to focus on the dense components (Figure \ref{fig:dlrma_pretraining}).}

Since embedding tables of DLRM-A make up 99.96\% of its 793B parameters, the only parallelization strategy viable for DLRM embedding tables on current GPU systems is naive model parallelism sharding.
This leaves parallelization strategy exploration on the base dense layers.
Figure~\ref{fig:dlrma_pretraining} demonstrates that, over valid parallelization strategies of the base dense layers on the x-axis, training throughput performance of DLRM-A can vary significantly from 0.19 (\texttt{(TP), (MP)}) to 1.14 $\times$ (\texttt{(TP, DDP), (MP)}) over the FSDP baseline.
Applying TP scales communication requirements with size of partial sums and activations.
If we apply TP at the intra-node level -- as opposed to globally -- we can take full of advantage of high BW NVLink to communicate the partial sums and activations.
In this case, since (\texttt{(DDP), (MP)}) replicates the dense layers' model parameters, gradients, and optimizer states across all devices, causes out-of-memory errors (OOM). 

\textbf{\underline{Insight 2: [LLMs]}
The billion-parameter scale of transformer layers in LLMs makes intra-node replication for compute layers infeasible.
In contrast, the small memory footprint of word embeddings (\textless 2GB) allows it to be replicated across all devices via DDP.}

In contrast to DLRMs, for LLMs (e.g., GPT-3), the FSDP offers competitive baseline training throughput (Figure~\ref{fig:overall_results}).
Since the word embeddings of LLMs are relatively small (0.37\% of GPT-3), full per-device embedding replication is a viable option via DDP.
As in the DLRM cases, we focus our parallelization strategy exploration on the compute-bound layers.
However, in the case of GPT-3, any form of layer replication across nodes (e.g., $\texttt{(TP, DDP)}$) leads to OOM since intra-node sharding is insufficient for meeting memory capacity requirements.
Additional device memory capacity can unlock up to 1.68$\times$ training throughput improvement.

\begin{figure}[t!]
    \centering
    \includegraphics[width=\linewidth]{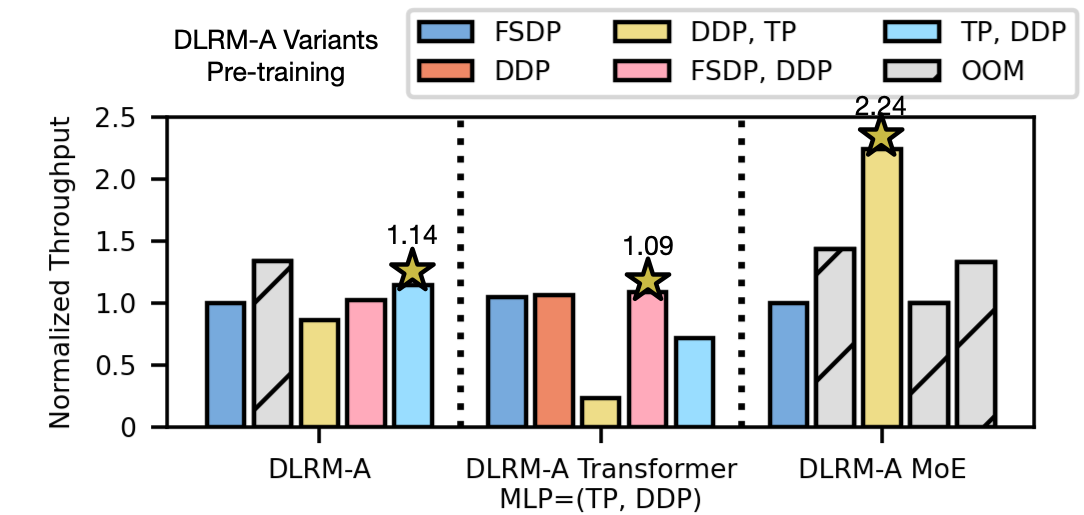}
    \caption{Between DLRM variants, both optimal parallelization strategy and expected throughput improvement vary.}
    \label{fig:dlrma_variants}
\end{figure}

\textbf{\underline{Insight 3: [Parallelization Strategy Order]}
Ordering of hierarchical parallelization strategies matter.
Replication and sharding strategies must be placed in the correct order to ensure optimal performance.
(Figures \ref{fig:overall_results}, \ref{fig:dlrma_pretraining}).}

The ``order" in which we apply hierarchical parallelization strategies matters greatly in terms of both memory capacity footprint and expected throughput.
For example, applying (\texttt{(TP), (DDP)}) shards the model component by \textit{number of devices in a node} while applying (\texttt{(DDP), (TP)}) shards the component by \textit{number of nodes}.
In Figure \ref{fig:dlrma_pretraining}, where there are 8 GPUs within a node and 16 nodes, the latter strategy leads to a lower per-GPU memory footprint.
Additionally, training throughput also varies from using different interconnect channels for communication.
For example, (\texttt{(TP), (DDP)}) leads to \texttt{AllReduce} of activations over faster NVLink and weight gradients over slower RoCE/IB.
On the other hand, (\texttt{(DDP), (TP)}) leads to communicating activations over RoCE/IB and weight gradients over NVLink.
For LLMs, long context lengths increase the size of activations to be communicated, so applying inter-node TP leads to significant slowdown (0.18$\times$ for GPT-3). 
On the other hand, utilizing NVLink to communicate large activations leads to 1.34$\times$ speedup for GPT-3.

\begin{figure}[t!]
    \centering
    \includegraphics[width=\linewidth]{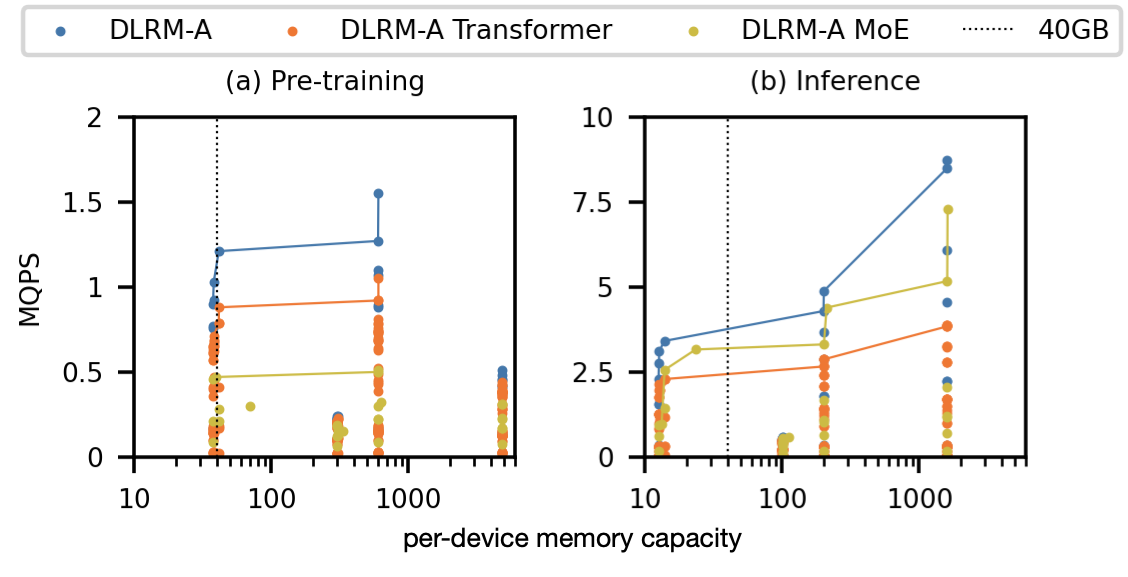}
    \caption{Pareto curves of parallelization strategies for DLRM variants for \textbf{(a)} pre-training and \textbf{(b)} inference.
    Each point is a different parallelization strategy.}
    \label{fig:variants_pareto}
\end{figure}

\begin{figure}[t!]
    \centering
    \includegraphics[width=\linewidth]{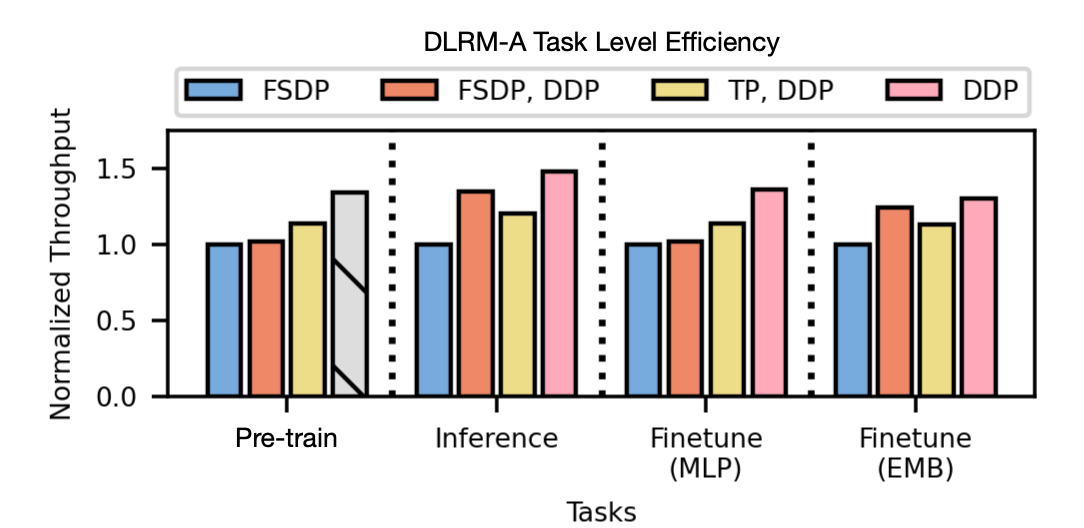}
    \caption{\srt{Task-level diversity (pre-training, inference, and fine-tuning) for the same underlying model and distributed system yields different amounts of speedup over FSDP baselines.}}
    \label{fig:dlrma_tasks}
\end{figure}

\begin{figure}[t!]
    \centering
    \includegraphics[width=\linewidth]{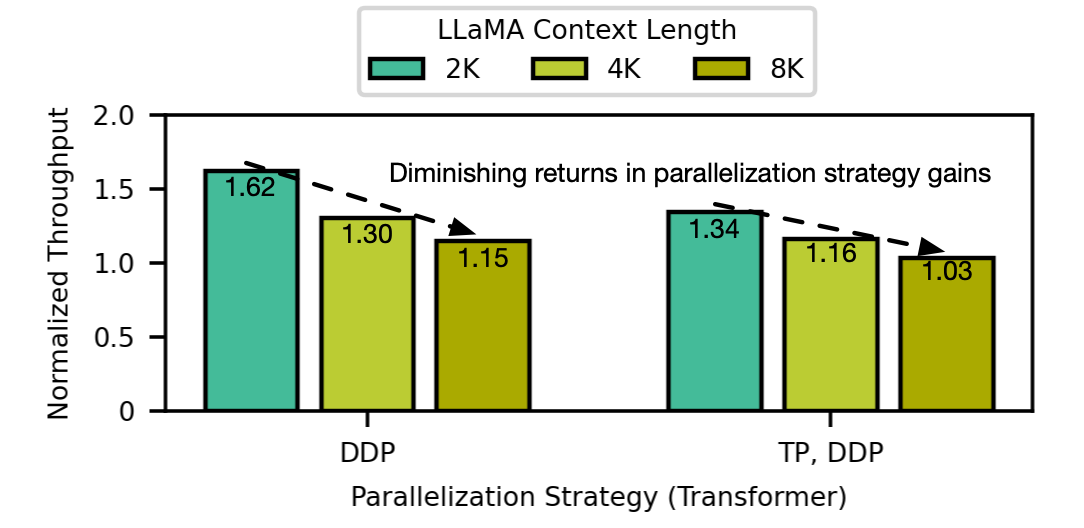}
    \caption{Given increasing context lengths, solely altering parallelization strategies has diminishing returns for performance benefits over FSDP.}
    \label{fig:contextlength}
\end{figure}

\textbf{\underline{Insight 4: [DLRM Variants]}
DLRM Transformer and MoE variants introduce new compute and communication requirements, leading to new parallelization strategy choice and task-level implications.
(Figures \ref{fig:dlrma_variants}, \ref{fig:variants_pareto}).}

Figure \ref{fig:dlrma_variants} shows how the same set of parallelization strategies interacts with both DLRM-A and its variants.
For DLRM-A Transformer, we apply (\texttt{(TP), (DDP)}) on the base dense layers since that is the optimal strategy for DLRM-A and focus parallelization strategy exploration on transformer layers.
Across the variants, optimal strategy (yellow star) varies.
These differences can be attributed to how transformers introduce more compute and more opportunities for communication-computation overlap while MoE increases blocking, non-overlapping All2All communication.
As models continue to evolve, parallelization strategies will as well.

\rewriteresults{
Figure~\ref{fig:variants_pareto} illustrates the parallelization strategy and model architecture options for DLRM-A, categorizing them by required per-device memory and potential throughput for pre-training and inference. 
The performance-pareto curve is marked with solid lines, indicating that higher memory capacity allows for strategies that achieve greater throughput. 
For pre-training, the transformer and MoE (Mixture of Experts) variants exhibit lower throughput due to increased computation and communication demands, respectively. 
During inference, the MoE variant shows greater efficiency compared to the transformer variant as the expensive expert communication is only necessary during the training's backward pass.
}

\newcontent{
\textbf{\underline{Insight 5: [Tasks]}
Inference, pre-training, and fine-tuning have different optimal parallelization strategies and scale-out efficiencies due to differences in forward and backward compute graphs (Figure \ref{fig:dlrma_tasks}).}
}

\srt{
Figure \ref{fig:dlrma_tasks} shows normalized DLRM-A throughput for various parallelization strategies in pre-training, inference, and fine-tuning.
For fine-tuning, we also evaluate the two different scenarios of fine-tuning MLP layers and embedding tables.
}

\srt{We see that certain parallelization strategies like DDP may be invalid for pre-training due to their excessive memory footprint requirements from storing per-device replicated model parameters, gradients, and optimizer states.
On the contrary, DDP becomes a viable option for inference and fine-tuning since memory footprint requirements are centered around parameters only for inference and parameters with subsets of gradients and optimizer states for fine-tuning.
The amount of speedup over FSDP baseline also varies for the different tasks.
Fine-tuning exclusively the embedding tables leaves less room for throughput improvement from different MLP sharding strategies.
Perhaps counter-intuitively, throughput-optimal parallelization strategy ordering for fine-tuning only embedding tables resembles that for inference.
This is because in this scenario we omit the costly MLP weight and input gradient calculations that are found during pre-training.}

\textbf{\underline{Insight 6: [Context-Length]}
Increasing context-lengths limits the improvements from parallelization strategy optimizations, necessitating either changes in model architecture or underlying distributed systems (Figure \ref{fig:contextlength}).}

Figure~\ref{fig:contextlength} shows that input complexity, in terms of context length, plays a key role in training throughput.
We investigate the effectiveness of (\texttt{(DDP)}) and (\texttt{(TP), (DDP)}) across LLMs of increasing context lengths.
2K and 4K context length examples refer to LLaMA and LLaMA2 while the 8K context length data point comes from doubling base LLaMA2's context length while keeping model architecture constant.

We see that throughput gains from tuning parallelization strategy decreases with increasing context length, indicating the limits of optimizing this design space.
To further improve throughput performance, changes have to be made to either the underlying distributed system or ML model architecture.

\begin{figure}[t!]
    \centering
    \includegraphics[width=\linewidth]{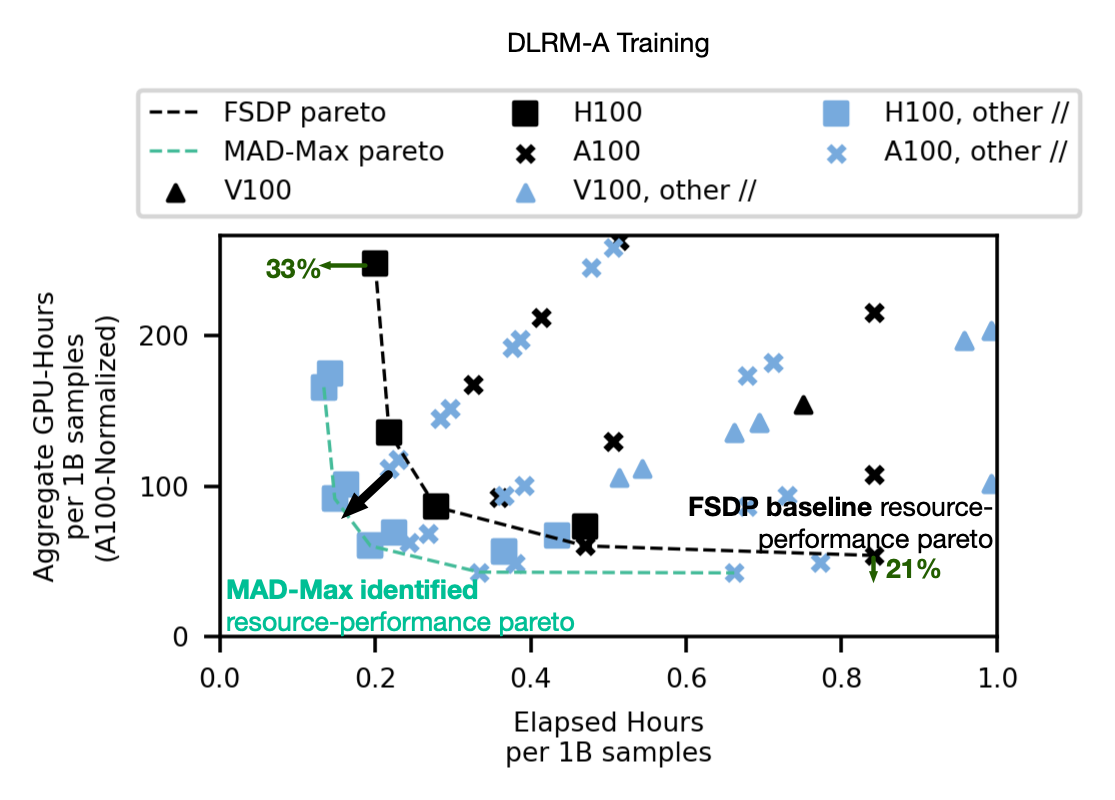}
    \caption{\srt{Across cloud instances of different GPUs and interconnects, parallelization strategy optimizations improve upon the resource-performance pareto frontier of FSDP baseline.
    Performance is quantified with elapsed time (hr) and compute resource requirements are quantified with aggregate GPU hours (normalized to A100 peak FLOPS).}}
    \label{fig:cloud}
\end{figure}

\textbf{\underline{Insight 7: [Cloud Deployment]} 
Optimizing cloud instance configurations and workload mappings improves both workload performance and operational compute resource requirements.}

\srt{Figure~\ref{fig:cloud} shows the training time (observed, elapsed hours) and compute resource requirements (aggregate GPU-hours, normalized to A100 peak FLOPS) of training DLRM-A across different GPU-instances from major public cloud providers.
To normalize aggregate GPU-hours across different generations of GPUs, we take each experiment's raw aggregate GPU-hours and normalize that number by the ratio between the target accelerator's peak FLOPS and A100 peak FLOPS.
This normalization is important for reflecting more accurate compute resource requirements since equal amounts of raw, aggregate GPU-hours between two clusters of different compute capabilities should correspond to different levels of resource requirements.
For those interested in exploring the trade-off space for other operational metrics, aggregate GPU-hours can also be potentially converted via other metric-specific ratios.
In this example, both performance and resource metrics correspond to processing 1 billion samples -- corresponding results for larger workloads (i.e., processing more samples) can be extrapolated using these ``per-1B samples" metrics.}

\srt{The pareto-optimal frontier established from using default FSDP parallelization strategies (black, dotted) can be improved upon by concurrently exploring different instance configurations (number of GPUs, networking capabilities) with parallelization strategies (green).
As seen in Figure~\ref{fig:cloud}'s legend, we include three generations of training-class NVIDIA GPUs, ranging from V100s to H100s.
For both V100 and A100 instances, both intra- and inter-node interconnect bandwidths vary greatly, with per-device inter-node interconnect bandwidths ranging from $<$1 to 25GB/s depending on the underlying RoCE or Infiniband specifications.
For intra-node interconnect, NVLink-enabled instances provide state-of-the-art performance.
For this DLRM-A training case study, we see up to 33\% training time and 21\% compute resource reduction.
By extension, operational energy consumption is also reduced due to less compute resources required -- as measured by aggregate GPU-hours -- for the task at hand.
Even if one were to explore this design space via an intelligent, constrained search, having a first-order performance model like MAD-Max for design guidance can enable aggregate GPU-hours savings on the order of 100s per 1 billion samples.
}

\input{tables/other_hw}

\begin{figure}[t!]
    \centering
    \includegraphics[width=\linewidth]{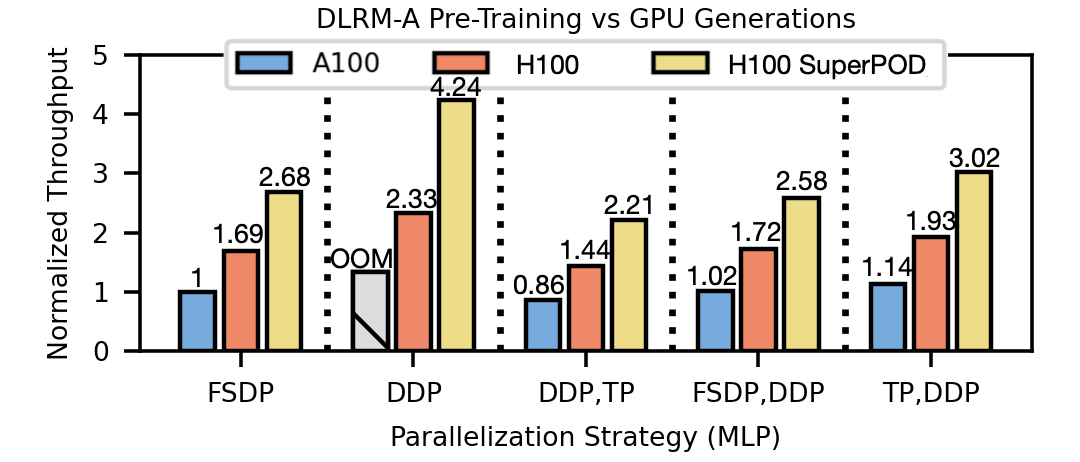}
    \caption{For DLRM-A pre-training, both overall GPU improvement (H100) and specifically upgrading inter-node interconnect fabric (H100 SuperPOD) lead to observable performance benefits.}
    \label{fig:gpu_generations}
\end{figure}

\textbf{\underline{Insight 8: [GPU-Generations]}
Across generations of GPUs, improvements in compute, memory, and interconnect not only improve distributed ML performance but also unlock different viable parallelization strategies.}

In Figure \ref{fig:gpu_generations}, we compare the A100 against a GPU with H100's specifications via simulation.
We also consider the H100 SuperPOD configuration, where the RoCE/IB inter-node interconnect fabric is replaced with NVLink for up to 256 GPUs, leading to $\sim$4.5$\times$ inter-node interconnect bandwidth compared to H100 DGX systems (see Table \ref{tbl:other_hw} for full specifications).

\rewriteresults{
Switching from the A100 (blue) to the H100 (orange) results in different levels of performance improvement across various parallelization methods. 
This variation in speedup is because the enhancements in compute, memory, and networking do not occur at the same rate when upgrading from A100 to H100. 
Additionally, each parallelization strategy prioritizes different aspects of the system's resources. 
Specifically, for DLRM-A training, solely upgrading the inter-node bandwidth (i.e., H100 to H100 SuperPOD) results in 1.82$\times$ higher throughput. 
This is primarily because the inter-node interconnect upgrade directly accelerates the blocking All2All embedding communication collectives.
}

\begin{figure}[t!]
    \centering
    \includegraphics[width=\linewidth]{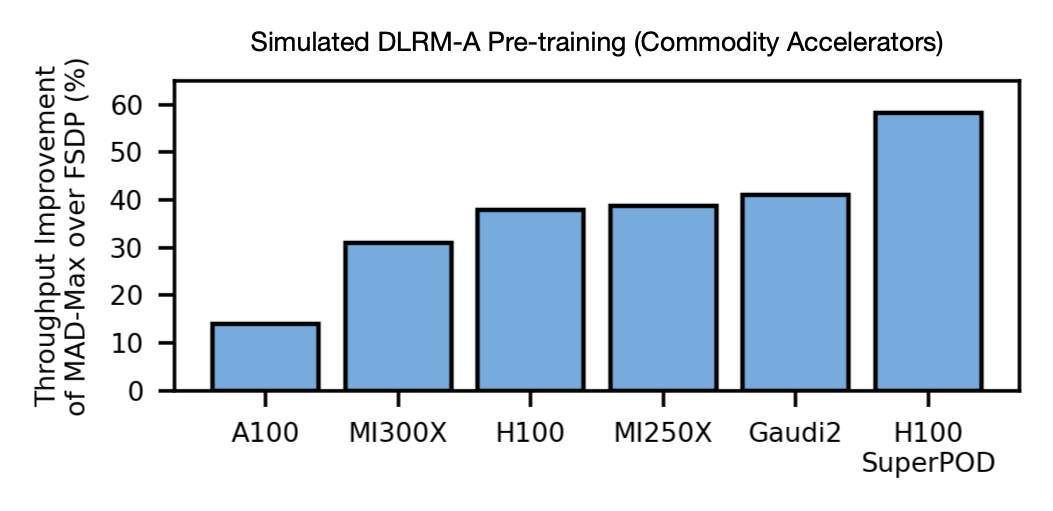}
    \caption{\srt{MAD-Max can simulate other commodity hardware (Table~\ref{tbl:other_hw}) and identify parallelization strategies that improve upon baseline FSDP.}}
    \label{fig:dlrm_a_newaccels}
\end{figure}

\newcontent{
\textbf{\underline{Insight 9: [Alternative Commodity Hardware]}
MAD-Max can simulate other commodity hardware platforms with independent compute and communication streams and further identify parallelization strategies with potential performance improvements.}
}

\srt{Figure~\ref{fig:dlrm_a_newaccels} depicts additional simulations for hardware configurations adjusted to best match AMD MI250X, MI300X GPUs and Intel Gaudi2 accelerators (Table \ref{tbl:other_hw}).
Similar to our baseline A100 ZionEX system~\cite{mudigere2021zionex}, we evaluate clusters of 128 devices for the DLRM-A pre-training task.
For AMD MI GPUs~\cite{amd2021mi250, amd2023mi300}, we follow reference scale-out CDNA platform designs~\cite{amd2021cdna2, amd2021cdna3}.
Since Gaudi2~\cite{intel2023gaudi2} does not have public datasheets, we follow prior benchmarking efforts on Intel Developer Cloud~\cite{databricks2024gaudi}.
We show results for throughput improvement from using a MAD-Max identified parallelization strategy over FSDP.
Compared to the 40GB-HBM A100, the other hardware platforms' increased HBM capacities (80+ GB) allow MAD-Max to identify parallelization strategies that replicate more dense model components for higher pre-training throughput.}

\begin{figure}[t!]
    \centering
    \includegraphics[width=\linewidth]{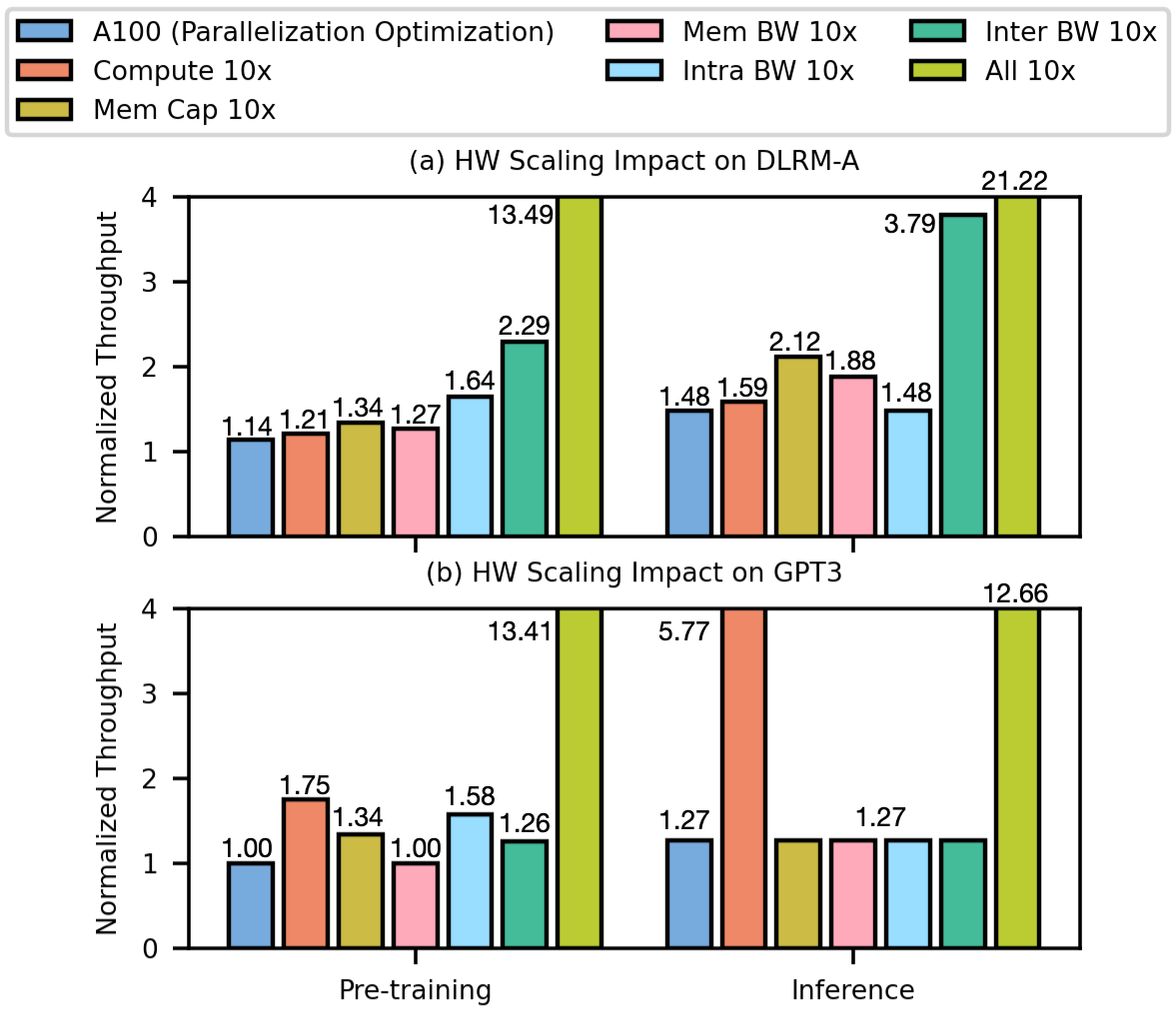}
    \caption{Individually scaling different hardware capabilities for \textbf{(a)} DLRM-A and \textbf{(b)} GPT-3 workloads leads to sub-linear speedup.
    Concurrently improving all capabilities leads to super-linear speedup.}
    \label{fig:hw_scaling}
\end{figure}

\begin{figure*}[t!]
    \centering
    \includegraphics[width=\linewidth]{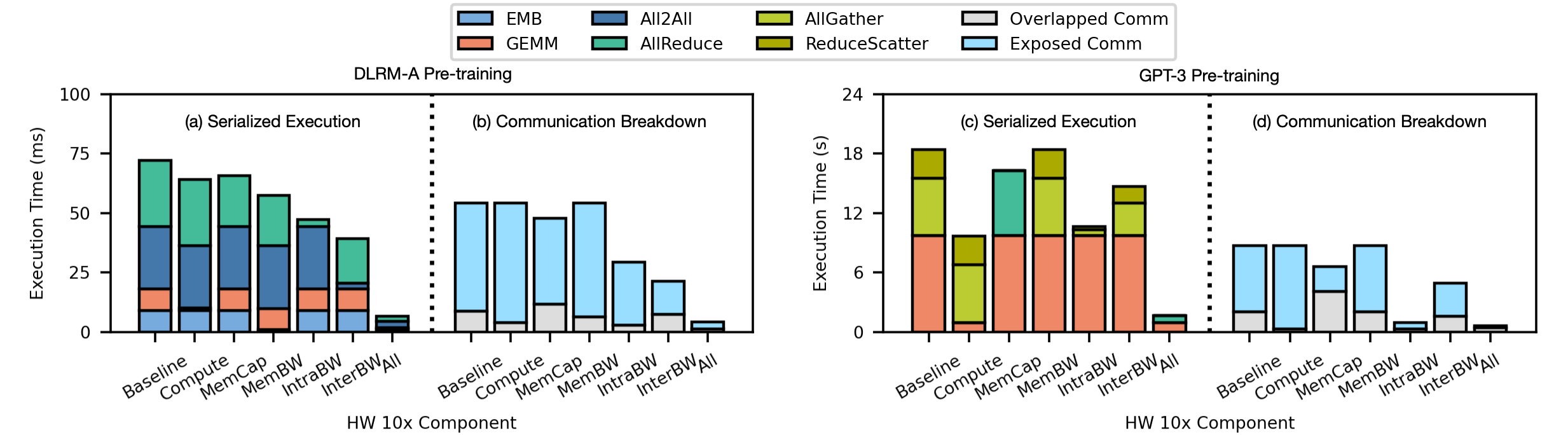}
    \caption{\textbf{(a, c)} Serialized execution and \textbf{(b, d)} communication breakdown for both DLRM-A and GPT-3 training allows us to better understand where speedups from hardware components come from.}
    \label{fig:hw_scaling_breakdown}
\end{figure*}

\textbf{\underline{Insight 10: [Future Technologies Trends]}
For large ML workloads, improving individual hardware components leads to limited throughput gain.
Unlocking further performance requires 
jointly improving hardware and systems specifications (Figures \ref{fig:hw_scaling}, \ref{fig:hw_scaling_breakdown}).}

From A100 to H100, compute, memory capacity, memory bandwidth, intra-node interconnect bandwidth, inter-node interconnect bandwidth improve by 2.42$\times$, 2$\times$, 1.29$\times$, 1.5$\times$, 2$\times$ (9$\times$ for SuperPOD), respectively.
In Figure \ref{fig:hw_scaling}, we perform a hardware scaling study where compute, memory capacity and bandwidth, intra- and inter-node interconnect bandwidth are all improved by 10$\times$ separately and concurrently.
We observe the effects of these improvements on DLRM-A and GPT-3 training and inference.

For DLRM-A pre-training and inference, independently improving anything but inter-node interconnect by 10$\times$ will only net 1.64 and 2.12$\times$ throughput improvements, respectively.
For these use-cases, since blocking All2All embedding communication is performance-critical, targeting inter-node communication bandwidth leads to substantial performance improvement.
For GPT-3, since compute-bound layers are critical to overall throughput, improving just compute throughput leads to more workload acceleration compared to DLRMs.

Figure \ref{fig:hw_scaling_breakdown} details the sources of the performance changes. 
Serialized execution breakdown shows execution time allocated to embedding lookups, GEMM, and specific communication collectives, disregarding the effects of overlap.
Computation-communication overlap breakdown shows how much communication is hidden behind embedding lookups and GEMM.
These breakdowns help us better understand the speedup results from Figure \ref{fig:hw_scaling} since throughput improvements can come from a variety of sources: accelerating compute-heavy layers (e.g., compute in GPT-3), reducing overall communication time (e.g., All2All in recommendation models), or even unlocking new parallelization strategies with more memory capacity (e.g., DDP for GPT-3).

For all four cases, jointly improving hardware components leads to super-linear performance improvement.
This is because distributed ML execution is non-serial so improving the performance of each trace segment can lead to more overlap or unlock new parallelization strategies altogether.

%% file: tables/other_hw.tex
\begin{table}[t!]
\begin{center}\resizebox{\linewidth}{!}{
\begin{tabular}{|c||c|c|c|c|}
\hline
\textbf{}                                                        & \textbf{FP-16/32 FLOPS} & \textbf{HBM Capacity, BW} & \textbf{\begin{tabular}[c]{@{}c@{}}Intra-Node BW\\ (per-device)\end{tabular}} & \textbf{\begin{tabular}[c]{@{}c@{}}Inter-Node BW\\ (per device)\end{tabular}} \\ \hline
\textbf{A100~\cite{nvidia2021a100}}                                                    & 312, 156 TFLOPS         & 40GB, 1.6TB/s             & 600GB/s                                                                       & 200Gbps                                                                       \\ \hline
\textbf{H100~\cite{nvidia2023h100}}                                                    & 756, 378 TFLOPS         & 80GB, 2TB/s               & 900GB/s                                                                       & 400Gbps                                                                       \\ \hline
\textbf{\begin{tabular}[c]{@{}c@{}}H100\\ SuperPOD~\cite{nvidia2023hgxh100}\end{tabular}} & 756, 378 TFLOPS         & 80GB, 2TB/s               & 900GB/s                                                                       & 1.8TBps                                                                       \\ \hline
\textbf{MI250X~\cite{amd2021mi250}}                                                  & 383, 96 TFLOPS          & 128GB, 3.2TB/s            & 500GB/s                                                                       & 200Gbps                                                                       \\ \hline
\textbf{MI300X~\cite{amd2023mi300}}                                                  & 1307, 654 TFLOPS        & 192GB, 5.3TB/s            & 896GB/s                                                                       & 400GBps                                                                       \\ \hline
\textbf{Gaudi2~\cite{intel2023gaudi2}}                                                  & 400, 200 TFLOPS         & 96GB, 2.5TB/s             & 262.5GB/s                                                                     & 300GBps                                                                       \\ \hline
\end{tabular}}
\end{center}
\caption{\srt{Simulated Commodity Hardware Specifications.}}
\label{tbl:other_hw}
\end{table}

%% file: text_CR/related_work_and_discussion.tex

\section{Related Work}~\label{sec:related_work}
We discuss related work in two key categories: parallelization strategy and distributed AI performance modeling 

\textbf{Parallelization Strategy Exploration.}
\cite{lepikhin2020gshard, xu2021gspmd} provide compiler annotations for identifying efficient parallelization strategies.
\cite{mahajan2023syndicate, rashidi2022themis} focus on optimizing communication collectives via fusion and scheduling.
\cite{zheng2022alpa} focuses on operator-level parallelism.
\cite{ardalani2022deepflow, jia2018flexflow} focus on parallelization strategy exploration but are evaluated on older and smaller ML models in Computer Vision and NLP.
\cite{zhang2017poseidon} explores strategies to overlap compute and communication before PyTorch.
In this paper, we aim to detach parallelization strategy exploration from existing software implementation details to enable an agile design space exploration of potentially yet to be implemented models.
Additionally, we target latest trillion-parameters scale models and expand our design space beyond just collectives.

\textbf{Distributed AI Performance Modeling.}
\cite{pope2022efficiently} provides an analytical model for transformer inference on TPUs.
\cite{pati2023computation} projects computation-communication overlap opportunities for future GPU-centric hardware.
\cite{rashidi2020astrasim, won2023astrasim2} provide a simulator for estimating distributed ML performance that is validated against AllReduce collectives.
\cite{kadiyala2022comet} builds upon \cite{rashidi2020astrasim, won2023astrasim2} to introduce a design space exploration tool, yet doesn't focus on optimizing training throughput for specific use cases like DLRM models.
These works build upon earlier work in simulating \cite{sanchez2013zsim, alian2017distgem5} and characterizing \cite{arpan2019perfchara, jeon2019multitenant} distributed systems.
\cite{wang2022topoopt} emphasizes network optimization.
\cite{liang2023mystique} focuses on generating replayable traces to better estimate hardware resource utilization.
\cite{sridharan2023chakra} is an effort to standardize traces across different software implementations for fair comparisons and generating synthetic traces, which can potentially be integrated with our performance model for better integration with current software implementations.
We design our performance model to be compatible with different hardware platforms, tasks, and exploration objectives.
We also focus on large ML model execution behavior and validate accordingly.

%% file: text_CR/conclusion.tex
\section{Conclusion}~\label{sec:conclusion}
Training and serving large-scale ML models is a resource-intensive and costly endeavor.
We present an agile performance modeling framework to identify efficient solutions for large-scale ML pre-training, fine-tuning, and inference that is also validated against large-scale experiments.
Using a suite of real-world large ML models, we identify parallelization strategies for improving performance on existing systems and cloud instances and performance bottlenecks of future hardware systems.

%% file: text_CR/acknowledgements.tex
\section*{Acknowledgements}
We thank Apostolos Kokolis, Giri Anantharaman, Kalyan Saladi and Srinivas Sridharan for feedback on modeling effective interconnect communication at-scale.
We also thank Can Balioglu, Changhan Wang, Kaushik Ram Sadagopan, and Yejin Lee for discussions on performance modeling and optimizations for key deep learning applications.